\def\spose#1{\hbox to 0pt{#1\hss}}
\def\ltsimm{\mathrel{\spose{\lower 3pt\hbox{$\sim$}}
        \raise 2.0pt\hbox{$<$}}}
\def\gtsimm{\mathrel{\spose{\lower 3pt\hbox{$\sim$}}
        \raise 2.0pt\hbox{$>$}}}
\def\Mdot{\hbox{${\dot M}$}}

\def\km{{\rm\thinspace km}}
\def\cm{{\rm\thinspace cm}}

\def\s{{\rm\thinspace s}}
\def\yr{{\rm\thinspace yr}}
\def\g{{\rm\thinspace g}}
\def\kmps{\hbox{${\rm\km\s^{-1}\,}$}}

\def\erg{{\rm\thinspace erg}}

\def\Hz{{\rm\thinspace Hz}}

\def\ster{{\rm\thinspace ster}}
\def\ergps{\hbox{${\rm\erg\s^{-1}\,}$}}

\def\Msol{\hbox{${\rm\thinspace M_{\odot}}$}}

\def\Msolpyr{\hbox{${\rm\Msol\yr^{-1}\,}$}}
\def\pcm{\hbox{${\rm\cm^{-1}\,}$}}
\def\pcm2{\hbox{${\rm\cm^{-2}\,}$}}
\def\pcm3{\hbox{${\rm\cm^{-3}\,}$}}
\def\ergpscm3Hz{\hbox{${\rm\ergps\cm^{-3}\Hz^{-1}\,}$}}
\def\ergpscm3Hzster{\hbox{${\rm\ergps\cm^{-3}\Hz^{-1}\ster^{-1}\,}$}}
\def\gpcm3{\hbox{${\rm\g\cm^{-3}\,}$}}
\def\ergpcm2{\hbox{${\rm\erg\cm^{-2}\,}$}}
\def\ergpcm3{\hbox{${\rm\erg\cm^{-3}\,}$}}
\def\phpscm2{\hbox{${\rm photons\s^{-1}\cm^{-2}\,}$}}

\documentclass[useAMS,usenatbib]{mn2e}
\usepackage{graphicx}
\usepackage{multirow}
\usepackage{amssymb}
\usepackage{epsfig,color}

\title[Self-Sealing Shells]{Self-Sealing Shells: Blowouts and Blisters on the Surfaces of
  Leaky Wind-Blown-Bubbles and Supernova Remnants}
\author[J.M. Pittard]{J.M. Pittard\\
School of Physics and Astronomy, The University of Leeds, Leeds, LS2 9JT}

\date{Accepted 2013 August 15.  Received 2013 August 13; in original form 2013 July 9}

\pagerange{\pageref{firstpage}--\pageref{lastpage}} \pubyear{2012}

\def\LaTeX{L\kern-.36em\raise.3ex\hbox{a}\kern-.15em
    T\kern-.1667em\lower.7ex\hbox{E}\kern-.125emX}

\begin{document}

\label{firstpage}

\maketitle

\begin{abstract}
  Blowouts can occur when a dense shell confining hot, high pressure,
  gas ruptures. The venting gas inflates a blister on the surface of
  the shell. Here we examine the growth of such blisters on the
  surfaces of wind-blown-bubbles (WBBs) and supernova remnants (SNRs)
  due to shell rupture caused by the Vishniac instability. On WBBs the
  maximum relative size of the blister ($R_{\rm bstall}/R$) is found
  to grow linearly with time, but in many cases the blister radius
  will not exceed 20 per cent of the bubble radius. Thus blowouts
  initiated by the Vishniac instability are unlikely to have a major
  effect on the global dynamics and properties of the bubble. The
  relative size of blisters on SNRs is even smaller than on WBBs, with
  blisters only growing to a radius comparable to the thickness of the
  cold shell of SNRs. The small size of the SNR blowouts is, however,
  in good agreement with observations of blisters in the Vela SNR. The
  difference in relative size between WBB and SNR blisters is due to
  the much higher speed at which gas vents out of WBBs, which
  translates into a greater energy flux through a rupture of a given
  size from interior gas of a given pressure. Larger blisters are
  possible if shell ruptures are bigger than expected.

  We expect the observed velocity structure of SNR shells to be
  affected by the presence of blisters until the shell is no longer
  susceptible to ruptures, since the initial expansion of blisters is
  faster than the ongoing expansion of the shell.
\end{abstract}

\begin{keywords}
hydrodynamics -- shock waves -- stars: mass loss -- ISM: bubbles --
stars: winds, outflows -- (ISM:) supernova remnants
\end{keywords}

%%%%%%%%%%%%%%%%%%%%%%%%%%%%%%%%%%%%%%%%%%%%%%%%%%%%%%%%%%%%%%%%%%%%%%%%%%%%%%%%
\section{Introduction}
The supersonic injection of material by stellar winds and/or
supernovae into the surrounding environment leads to the formation of
wind-blown bubbles (WBBs) and supernova remnants (SNRs). The gas swept
up by these structures usually cools to form a dense shell which may
be prone to the Vishniac \citep{Vishniac:1983} overstability.  This
instability can grow from a linear perturbation and is characterised
by an oscillation of the shock front and a power-law growth rate.

If the shell ruptures, hot, rarefied gas should flow at high speed
into the undisturbed environment. In a study of the Vela SNR,
\citet*{Meaburn:1988} concluded that such ruptures were responsible
for inflating ``blisters'' on the surface of the swept-up shell. Since
the gas swept-up by a blister also cools and forms a shell, the shell
effectively ``self-seals''. In the frame of reference of the main
shell the blister will expand to a certain radius, stall, and then
merge back into the shell. \citet{Meaburn:1988} claim that the
blisters that they detect are of comparable size to the thickness of
the shell.  The Vishniac instability has also been suggested to be
responsible for the filamentary structure seen in older SNRs
\citep[see][and references therein]{MacLow:1993} and in the Eastern
part of the WBB NGC~6888 \citep{Garcia-Segura:1996}.

In this work we study blisters on the surfaces of both WBBs and
SNRs. Though the necessary conditions for shell rupture are not well
known, we can nevertheless estimate the size of a resulting blister
should a shell rupture occur. First we derive an expression for the
evolution of the blister radius and determine the size of the blister
at the time that it stalls. We then determine how the maximum size of
a blister scales with the age of the underlying bubble or remnant, and
we determine the maximum size of the blister relative to the original
bubble/remnant. We discuss our analytical results for leaky WBBs in
relation to numerical simulations where the Vishniac instability has
been noted.  We also compare our analytical results for leaky SNRs
with observational data of remnants.

\section{Analytical Model}
\label{sec:analytical_model}

\subsection{General Considerations}
\label{sec:analytical_general}
The thickness of a cold WBB or SNR shell can be determined from mass
conservation to be
\begin{equation}
\label{eq:dR}
\delta R = \frac{R}{5 M^{2}},
\end{equation}
where $M$ is the {\em adiabatic} Mach number of the shell into the
surrounding medium and $R$ is its radius\footnote{This is equivalent
  to $\delta R = \frac{R}{3 M_{\rm iso}^{2}}$ where $M_{\rm iso}$ is
  the {\em isothermal} Mach number of the shell.}.

For expansion into a uniform environment the shell is continually
decelerating and so is stable against Rayleigh-Taylor (RT)
instabilities. However, the shell can be unstable to the Vishniac
instability. In such cases, the shell is most
unstable for perturbations with a wavenumber $k$ where $k\,\delta R \sim
1$. This corresponds to a wavelength $\lambda = 2 \pi/k \sim 2 \pi
\delta R$ and a spherical harmonic mode number $l = kR$.  The growth
timescale for the instability is
\begin{equation}
\label{eq:dtvish}
\delta t_{\rm vish} = \left(\frac{\sigma}{P k^{2} |\ddot{R}|}\right)^{1/4},
\end{equation}
where $\sigma$ is the surface density of the shell ($\sigma = R
\rho_{\rm 0}/3$), $P$ is the interior thermal pressure of the hot gas
and $|\ddot{R}|$ is the effective deceleration of the shell.  If the
instability ruptures the shell it will create a gap through which the
hot pressurized gas in the bubble/remnant interior can vent. The area
of the ``nozzle'' $A \approx \pi \lambda^{2}/4 \approx \pi^{3} \delta
R^{2}$. We assume that the size of the nozzle is constant in time,
though in reality $A$ is likely to increase due to the continued
expansion of the main bubble/remnant and due to the ablation/stripping
of material from its sides by the hot gas flowing through it. As well
as widening the nozzle, this latter process would ``mass-load'' the
interior of the blister.

The rate of energy release from the bubble or remnant interior into the
``blowout'' is given by the flux of kinetic and thermal energy through
the nozzle,
\begin{equation}
\label{eq:edotvent}
\dot{E}_{\rm v} = \left[\frac{1}{2}\rho v^{3} + \frac{P v}{\gamma -
    1}\right] A,
\end{equation}
where $\rho$ is the interior density of the hot gas and $v$ is the
flow speed through the nozzle. The flow through such a nozzle will be
transonic so we set $v = c$ where $c$ is the sound speed of the
interior hot gas.

The vented material flows supersonically into the undisturbed ambient
medium, and sweeps it up, so that a blister is formed on the surface
of the bubble/remnant. The blister is bounded by a cold shell once its
age exceeds the cooling time of the swept-up gas. The blister
initially expands relatively rapidly, but like the main bubble/remnant
it slows as it sweeps up more and more material. We assume that the
blister ``stalls'' once its expansion velocity drops below that of the
main bubble/remnant.  While the blister is relatively modest in size
(e.g. $R_{\rm b} \ltsimm 0.1\,R$ where $R_{\rm b}$ is the blister
radius), its shape is close to hemi-spherical. If it stalls while this
condition is true its evolution follows that of a standard
pressure-driven bubble \citep[e.g.][]{Dyson:1980}, but with a slightly
different constant of proportionality:
\begin{equation}
\label{eq:rblister}
R_{\rm b} = \left(\frac{125}{77 \pi}\right)^{1/5}
\left(\frac{\dot{E}_{\rm v}}{\rho_{\rm 0}}\right)^{1/5} t_{\rm b}^{3/5}.
\end{equation}
Note that the rate of energy venting, $\dot{E}_{\rm v}$, and the time since
the blowout, $t_{\rm b}$, replace the equivalent $\dot{E}$ and $t$ of
a standard WBB. 

The swept-up gas becomes thermally unstable and collapses into a thin
shell when the cooling time becomes less than the age of the
blister. Assuming a strong adiabatic shock, a cold shell forms at
\begin{equation}
\label{eq:tcool}
\delta t_{\rm cb} = (2.3\times10^{4}\,{\rm yr})\,n_{\rm
  0}^{-0.71}\,\dot{E}_{\rm v,38}^{0.29},
\end{equation}
where $n_{\rm 0}$ is the number density of the ambient medium and
$\dot{E}_{\rm v,38}$ is the vented energy flux in units of $10^{38}\,\ergps$ \citep{MacLow:1988}.
If the blister expands into ionized gas with $\mu {\rm m_{H}} \approx 10^{-24}$\,g, 
\begin{equation}
\delta t_{\rm cb} = 9.9\times10^{-17} \rho_{\rm 0}^{-0.42}
\dot{E}^{0.145} \dot{M}^{-0.145} c_{\rm 0}^{1.16} t^{0.58},
\end{equation}
where $c_{\rm 0}$ is the adiabatic sound speed of the ambient medium
and cgs units are used throughout.
  
The age of the blister when its expansion velocity drops below
that of the main bubble/remnant is
\begin{equation}
\label{eq:tstall}
\delta t_{\rm stall} = \left(\frac{125}{77 \pi}\right)^{1/2}
\left(\frac{\dot{E}_{\rm v}}{\rho_{\rm 0}}\right)^{1/2} \left(\frac{5
    \dot{R}}{3}\right)^{-5/2}.
\end{equation}
At this time, the radius of the blister is
\begin{equation}
\label{eq:rblisterstall}
R_{\rm bstall} = \left(\frac{125}{77 \pi}\right)^{1/5}
\left(\frac{\dot{E}_{\rm v}}{\rho_{\rm 0}}\right)^{1/5} \delta t_{\rm stall}^{3/5},
\end{equation}
and the ratio $R_{\rm b}/R$ is a maximum for this particular blister. After this time the
blister begins to merge back into the main bubble/remnant as it is overtaken
by material in the adjacent cold shell.
Substituting Eq.~\ref{eq:tstall} into Eq.~\ref{eq:rblisterstall} gives
\begin{equation}
\label{eq:rblisterstall1.5}
R_{\rm bstall} = \left(\frac{125}{77 \pi}\right)^{1/2}
\left(\frac{\dot{E}_{\rm v}}{\rho_{\rm 0}}\right)^{1/2} \left(\frac{5 \dot{R}}{3}\right)^{-3/2}.
\end{equation}

We assume that the main WBB/SNR shell is susceptible to the Vishniac
instability until its Mach number declines to the point where the
shell becomes too thick for the instability to grow. This occurs at
$t=t_{\rm nr}$ when the density jump at the shell corresponds to an
effective ratio of specific heats of $\gtsimm 1.1$ for WBBs
\citep{Ryu:1988} and $\gtsimm 1.2$ for SNRs
\citep{Ryu:1987,Vishniac:1989}.

\subsection{Wind-Blown Bubble Specifics}
\label{sec:analytical_wbbs}
Mass and energy injection by a star or group of stars is often
observed to inflate a bubble in the surrounding medium.
The swept-up gas becomes thermally unstable and collapses into a thin
shell when the cooling time becomes less than the age of the
bubble. The radius of a pressure-driven bubble with a thin swept-up shell is
\citep[e.g.][]{Dyson:1980}

\begin{equation}
\label{eq:rbub}
R = \left(\frac{125}{154 \pi}\right)^{1/5}
\left(\frac{\dot{E}}{\rho_{\rm 0}}\right)^{1/5} t^{3/5},
\end{equation}
where $\dot{E}$ is the rate of energy injection into the bubble by
stellar winds and/or supernovae, $\rho_{\rm 0}$ is the mass density of
the ambient medium, and $t$ is the age of the bubble. 

The thermal pressure within the bubble is given by 
\begin{equation}
\label{eq:pbub}
P = \rho_{\rm 0}\dot{R}^{2} + \frac{1}{3}\rho_{\rm 0}\ddot{R}R,
\end{equation}
where $\dot{R}$ and $\ddot{R}$ are the expansion speed and
acceleration of the bubble, respectively \citep{Dyson:1980}. Substituting for $\dot{R}$ and $\ddot{R}$ one obtains
\begin{equation}
\label{eq:pbub2}
P = \frac{7}{(3850\pi)^{2/5}} \dot{E}^{2/5} \rho_{0}^{3/5} t^{-4/5}.
\end{equation}
The average density within the bubble is
\begin{equation}
\label{eq:rhobub}
\rho \approx \frac{3\dot{M}t}{4 \pi R^{3}},
\end{equation}
where $\dot{M}$ is the rate of mass injection into the bubble by
  stellar winds and/or supernovae and we have ignored the volume
occupied by the freely streaming wind in the bubble centre. We have
also assumed that there is no evaporation of mass from the swept-up
shell into the bubble interior \citep{Weaver:1977}, or mass-loading
from the destruction of clouds overrun by the shell
\citep[e.g.][]{Pittard:2001a,Pittard:2001b}.  The interior sound speed
of the bubble is $c = (\gamma P/\rho)^{1/2}$. Substituting this, the
equation for $A$, and Eqs.~\ref{eq:pbub2} and~\ref{eq:rhobub} into
Eq.~\ref{eq:edotvent} gives, for $\gamma=5/3$,
\begin{equation}
\label{eq:edotvent_wbb}
\dot{E}_{\rm v} = 8.74\, \rho_{0}^{3/2} R^{5/2} \dot{M}^{-1/2}
c_{0}^{4} t^{1/2}.
\end{equation}
Since $R \propto t^{3/5}$, we find that $\dot{E}_{\rm v} \propto
t^{2}$.

Substituting Eq.~\ref{eq:edotvent_wbb} into
Eq.~\ref{eq:rblisterstall1.5}, and replacing $\dot{R}$ with $3R/5t$, yields
\begin{equation}
\label{eq:rblisterstall2}
R_{\rm bstall} = 2.27\,\rho_{\rm 0}^{3/10}\dot{M}^{-1/4}c_{\rm
  0}^{2}\dot{E}^{-1/20} t^{8/5}.
\end{equation}
Since the radius of the main bubble increases as $t^{3/5}$, we see
that the maximum relative size of a blister $(R_{\rm bstall}/R)$
increases linearly with the bubble age, $t$.

Since $\delta R = R/(5 M^{2})$ and $M = \dot{R}/c_{0} = 3 R/(5 c_{\rm
  0} t)$, the thickness of the swept-up shell increases with time as
\begin{equation}
\label{eq:drvst}
\delta R = 0.728\,c_{0}^{2} \left(\frac{\dot{E}}{\rho_{0}}\right)^{-1/5} t^{7/5}.
\end{equation}
Hence the relative thickness of the shell is
\begin{equation}
\label{eq:drR}
\frac{\delta R}{R} = 0.955\, c_{0}^{2}
\left(\frac{\dot{E}}{\rho_{0}}\right)^{-2/5} t^{4/5}.
\end{equation}
Combining Eqs.~\ref{eq:rblisterstall2} and~\ref{eq:drvst} gives the
ratio of the maximum size of a blister to the shell thickness as
\begin{equation}
\label{eq:rbstalldr}
\frac{R_{\rm bstall}}{\delta R} = 3.12 \rho_{0}^{1/10}\,\Mdot^{-1/4}\,\dot{E}^{3/20}t^{1/5}.
\end{equation}

As noted in Sec.~\ref{sec:analytical_general}, rupturing of the shell
by the Vishniac instability stops when the density jump at the shell
corresponds to an effective ratio of specific heats of $\gtsimm 1.1$.
The density jump at an adiabatic shock with $\gamma \approx 1.1$ is $\approx (\gamma +
1)/(\gamma - 1) \approx 21$. An isothermal shock produces an
equivalent density jump for an isothermal Mach number $M_{\rm iso}
\approx \sqrt{21}$. For a $\gamma = 5/3$ gas this corresponds to an
adiabatic Mach number $M = M_{\rm iso}/\sqrt{\gamma} \approx 3.55$.
Thus we assume that shell ruptures stop when the
shell Mach number drops below 3.55. This corresponds to a time of
\begin{equation}
\label{eq:t_novish}
t_{\rm nr} \approx 0.169\frac{R}{c_{\rm 0}} \approx 5.965\times10^{-3}
c_{\rm 0}^{-5/2} \left(\frac{\dot{E}}{\rho_{\rm 0}}\right)^{1/2}.
\end{equation}
Substituting into Eq.~\ref{eq:rblisterstall2} gives the maximum
possible size of any blister (which occurs when $t=t_{\rm nr}$) as
\begin{equation}
\label{eq:rblisterstall3}
R_{\rm bmax} = 6.3\times10^{-4}\,\rho_{\rm 0}^{-1/2}\dot{M}^{-1/4}c_{\rm
  0}^{-2}\dot{E}^{3/4}.
\end{equation}
The radius of the main bubble at this time is
\begin{equation}
\label{eq:rbub2}
R = 0.0353 \left(\frac{\dot{E}}{\rho_{\rm 0}}\right)^{1/2} c_{\rm
  0}^{-3/2}.
\end{equation}
Hence, the maximum relative size of the blister is
\begin{equation}
\label{eq:rblistermaxrel}
\frac{R_{\rm bmax}}{R} = 1.8\times10^{-2}\,\left(\frac{\dot{E}}{\dot{M}}\right)^{1/4}c_{\rm
  0}^{-1/2}.
\end{equation}
Writing $\dot{E} = 0.5\dot{M}v_{\rm wind}^{2}$, we finally obtain
\begin{equation}
\label{eq:rblistermaxrel2}
\frac{R_{\rm bmax}}{R}= 1.5\times10^{-2}\,\left(\frac{v_{\rm wind}}{c_{\rm 0}}\right)^{1/2}.
\end{equation}
At $t=t_{\rm nr}$ it also follows that
\begin{equation}
\label{eq:dr_tnr}
\delta R \approx 5.6 \times 10^{-4} c_{0}^{-3/2} \left(\frac{\dot{E}}{\rho_{0}}\right)^{1/2},
\end{equation}
and
\begin{equation}
\label{eq:rbmaxdr}
\frac{R_{\rm bmax}}{\delta R} = 1.126\,c_{\rm 0}^{-1/2}
\left(\frac{\dot{E}}{\Mdot}\right)^{1/4} = 0.947\,\left(\frac{v_{\rm wind}}{c_{0}}\right)^{1/2}.
\end{equation}

\subsection{Supernova Remnant Specifics}
\label{sec:analytical_snrs}
Supernova explosions drive high Mach number shocks into the
surrounding interstellar medium (ISM). The shock-heated gas forms a
SNR which initially expands freely, but which slows as the mass it
sweeps up becomes comparable to and then exceeds the ejecta mass. The
swept-up mass is concentrated in a thick shell behind the forward
shock and the whole remnant is hot ($T\gtsimm10^{7}$\,K) and expands
adiabatically in what is known as its Sedov-Taylor (ST) stage
\citep{Taylor:1950,Sedov:1959}. Eventually the expansion speed of the
SNR becomes low enough that the post-shock gas begins to lose
significant energy through radiative cooling. The thick shell becomes
compressed into a thin dense shell of cooled gas which continues to be
driven forward by the high pressure, hot and low-density interior in
what is known as the pressure-driven snowplough (PDS) stage
\citep{McKee:1977}.  It is at this time that the Vishniac instability
may play an important role.

The key input parameters for our model are the initial mechanical
energy of the supernova explosion, $E_{\rm 0}$, and the density and
temperature of the ambient medium, $\rho_{\rm 0}$ and $T_{\rm 0}$
respectively, which we assume are uniform.

The transition of the SNR to the radiative phase has been studied in
detail by a number of works including \citet{Blondin:1998}. Assuming a
strong shock, and an ionized medium with cosmic abundances, the
cooling time of gas immediately behind the forward shock of the SNR is
equal to the age of the remnant at the transition time
\begin{equation}
\label{eq:t_trans}
t_{\rm tr} \approx (2.9\times10^{4}\,{\rm yr})\,n_{\rm H0}^{-9/17}\,E_{51}^{4/17},
\end{equation}
where $n_{\rm H0}$ is the number density of hydrogen in the ambient
medium and $E_{51} = E_{\rm 0}/(10^{51}\,{\rm ergs})$. The radius and
velocity of the remnant at this time are respectively 
\begin{equation}
\label{eq:r_trans}
R_{\rm tr} \approx (19.1\,{\rm pc})\,n_{\rm H0}^{-7/17}\,E_{51}^{5/17}
\end{equation}
and
\begin{equation}
\label{eq:v_trans}
\dot{R}_{\rm tr} \approx (260\,\kmps)\,n_{\rm H0}^{2/17}\,E_{51}^{1/17}.
\end{equation}
\citet{Blondin:1998} note that the
actual time of shell formation, as determined from hydrodynamical
calculations, is $t_{\rm sf} \approx 1.6\times t_{\rm
  tr}$. The value of $t_{\rm sf}/t_{\rm tr}$ is
sensitive to the ambient density, increasing from 1.32 when $n_{\rm
  H0}=8400\,\pcm3$ to 1.85 for $n_{\rm H0}=0.084\,\pcm3$.

To calculate the growth timescale of the Vishniac instability we need
to know the rate of deceleration of the shell. The radius of remnants
is often described in terms of a power-law with respect to time:
$R\propto t^{m}$. During the ST stage $R_{\rm ST} \propto t^{2/5}$,
while in the pressure-driven snowplough stage $R_{\rm PDS} \propto
t^{2/7}$. However, simulations show that there is not a smooth
transition between the ST and PDS stages and that in practice thin
shell formation in SNRs often generates secondary shocks
\citep[e.g.,][]{Falle:1981}. These cause the forward shock and the
deceleration parameter, $m = d({\rm log}\, R)/d({\rm log}\, t)$, to
undergo strong oscillations for a period of time after formation of
the cold shell. Numerical studies show that after these oscillations
have damped out $m$ increases to a maximum value of 0.33
\citep[e.g.,][]{Blondin:1998}, before converging asymptotically to the
pressure-driven snowplough value of $2/7$
\citep{McKee:1977,Bandiera:2004}.

To avoid having to conduct full numerical simulations of this process
we make use of the simple functional fit to $m$ presented by
\citet{Bandiera:2004}.  For the slow branch (the physically relevant
case) the deceleration parameter
\begin{equation}
\label{eq:m_bandiera}
m_{\rm s}(r) = \frac{2}{35} \frac{r-1}{r^{4}}(5r^{3} + 6r^{2} + 8r
+ 16) + \frac{\sqrt{r-1}}{r^{4}} C,
\end{equation}
where $C=-0.248\pm0.006$ and $r = 1.176\, R/R_{\rm tr}$.
\citet{Bandiera:2004} also provide an expression for
$\tau=t/\tilde{t}$ as a function of $r$:
\begin{equation}
\label{eq:tau_bandiera}
\tau_{\rm s}(r) = \frac{2}{35}\sqrt{r-1}(5r^{3} + 6r^{2} + 8r + 16) + C,
\end{equation}
where $\tilde{t} = 1.14 \,t_{\rm tr}$. Hence for a specified value of
$r$ we can obtain the age of the remnant $t$ and also a value for the
deceleration parameter, $m$. Since $m = \dot{R}t/R$, where $\dot{R}$
is the expansion speed of the remnant, we can then determine the shell
velocity and adiabatic Mach number, $M$. The deceleration rate of the
shell is given by $\ddot{R} = (m-1)\dot{R}/t$.
  
For simplicity we assume that a thin shell has formed once $t\gtsimm
1.6\,t_{\rm tr}$, and that the mass swept-up by the remnant is fully
contained within it. This should be a good approximation since the
interior density of a SNR is very low\footnote{At the time of
  transition into the radiative phase $\approx 50$ per cent of the
  interior mass is compressed into the thin shell. However, the mass
  in the thin shell approaches the swept up mass at later times
  \citep{Chevalier:1974,Mansfield:1974}.}.

The pressure jump at the shock is given by the standard
Mach-number dependent Rankine-Hugoniot condition: 
\begin{equation}
\label{eq:p_postshock}
P_{\rm s} = \frac{2\gamma M^{2} - (\gamma-1)}{\gamma+1}\,P_{\rm 0}.
\end{equation}
where $P_{\rm 0}$ is the pressure of the ambient medium.

The exact values to use for $\rho$ and $P$ in Eq.~\ref{eq:edotvent}
are a little uncertain because the density and pressure of the gas
inside the remnant varies with the distance from the centre of the
remnant. For guidance we refer to the numerical results of
\citet{Blondin:1998} where they claim that only the outermost 5 per
cent of the remnant structure is affected by radiative
cooling\footnote{Note that the gas adjacent to the shell is also cool
  at the time of shell formation, though a reverse shock develops and
  heats gas in this region after $t=2.0-2.5\,t_{\rm tr}$ \citep[see
  Fig.~6 in][]{Blondin:1998}.}. We therefore make the approximation
that the properties of the flow through the nozzle are equivalent to
the hot phase at a radius of $\approx 0.93 R$. From Fig.~5 in
\citet{Blondin:1998} we then determine that $\rho \approx
1.5\,\rho_{\rm 0}$ and $P \approx 0.3 P_{\rm s}$.  The sound speed of
this gas then follows: $c = \sqrt{\gamma P/\rho}$.

If the cold shell expands at a sufficiently high Mach number,
Eq.~\ref{eq:p_postshock} can be approximated as
\begin{equation}
\label{eq:p_postshock_highM}
P_{\rm s} \approx \frac{2 \gamma M^{2}}{\gamma + 1}\,P_{\rm 0},
\end{equation}
and then
\begin{equation}
\label{eq:v_nozzle}
v \approx c \approx \sqrt{\frac{\gamma P}{\rho}} \approx \gamma
\sqrt{\frac{2 M^{2}}{5(\gamma+1)}\frac{P_{\rm 0}}{\rho_{\rm 0}}}.
\end{equation}
Eq.~\ref{eq:edotvent} then evaluates to 
\begin{equation}
\label{eq:edotvent2}
\dot{E}_{\rm v} \approx 0.701\, P_{\rm 0}^{3/2} \rho_{\rm 0}^{-1/2} R^{2} M^{-1}.
\end{equation}
Substituting Eq.~\ref{eq:edotvent2} into Eq.~\ref{eq:rblister} yields:
\begin{equation}
\label{eq:rblistergrow}
R_{\rm b} \approx 0.7 m^{-1/5} c_{\rm 0}^{4/5} R^{1/5} t_{\rm b}^{3/5}
t^{1/5}.
\end{equation}
Since the remnant radius $R= a\,t^{m}$, the radius of an individual blister increases
with time as $R_{\rm b} \propto m^{-1/5} t^{(1+m)/5} t_{\rm b}^{3/5}$ until it reaches
its maximum radius (at $t_{\rm b} = \delta t_{\rm stall}$):
\begin{equation}
\label{eq:rblisterattstall}
R_{\rm bstall} \approx 0.191\,m^{-2} c_{\rm 0}^{2} R^{-1} t^{2}.
\end{equation}
Thus the maximum blister size as a function of the
remnant age, $t$, grows as $R_{\rm bstall} \propto m^{-2} t^{2-m}$.
The relative size of such blisters is
\begin{equation}
\label{eq:relsize_vs_t}
\frac{R_{\rm bstall}}{R} \approx 0.191\left(\frac{c_{\rm 0}}{m
    \,a}\right)^{2} t^{2-2m}.
\end{equation}
Since 
\begin{equation}
\delta R = \frac{R}{5 M^{2}} = \frac{1}{5R}\left(\frac{c_{\rm
      0}t}{m}\right)^{2},
\end{equation}
we further find that
\begin{equation}
\label{eq:rblister_over_dR}
\frac{R_{\rm bstall}}{\delta R} \approx 0.954.
\end{equation}
Hence individual blisters do not grow larger than the thickness of the
shell! This is in stark contrast to the case of blisters formed by
the shell ruptures of wind-blown-bubbles, which as we have seen in
Sec.~\ref{sec:analytical_wbbs} can grow much
larger than the shell thickness.

The Vishniac instability stops at $t = t_{\rm nr}$ once the cold shell
becomes too thick, which as noted in Sec.~\ref{sec:analytical_general}
occurs when the density jump at the shell corresponds to an effective
ratio of specific heats of $\gtsimm 1.2$. Using the same argument as
in Sec.~\ref{sec:analytical_wbbs}, we find that this corresponds to an
adiabatic Mach number $M \approx 2.57$ for SNRs i.e. when $\dot{R}
\approx 2.57\,c_{\rm 0}$. At this time $m$ is near its maximum value
of 0.33 \citep[see Fig.~1 of][]{Bandiera:2004}. Since $\dot{R} = m
R/t$, and $r=1.176\,R/r_{\rm tr}$, we find that
\begin{equation}
\label{eq:t_nr}
t_{\rm nr} = m \frac{R}{\dot{R}} \approx \frac{0.33}{2.57} \frac{r}{c_{\rm
    0}} \approx 0.128\,\frac{R}{c_{\rm 0}}. 
\end{equation} 
This compares to $t_{\rm nr} \approx 0.169\,R/c_{\rm 0}$ for leaky wind-blown
bubbles.

At $t=t_{\rm nr}$, $\delta R/R \approx 1/33$ and $P_{\rm s} \approx
8\,P_{\rm 0}$, so $P = 12\,P_{\rm 0}/5$. The velocity of gas through
the nozzle is then $v = \sqrt{8\,P_{\rm 0}/(3\,\rho_{\rm 0})}$. With
$A = \pi^{3}\,{\rm \delta R^{2}}$ we obtain
\begin{equation}
\label{eq:e_vent_tnr}
\dot{E}_{\rm v} \approx 0.26\,P_{\rm 0}^{3/2} \rho_{\rm 0}^{-1/2} R^{2},
\end{equation}
so that the maximum possible size of a blister on a remnant shell is 
\begin{equation}
\label{eq:rblisternr}
R_{\rm bmax} = \left(\frac{125}{77 \pi}\right)^{1/2} \left(\frac{3}{5}\right)^{3/2}
\left(\frac{\dot{E}_{\rm v}}{\rho_{\rm 0}}\right)^{1/2} \left(\dot{R}_{\rm
  nr}\right)^{-3/2},
\end{equation}
where $\dot{R}_{\rm nr}$ is the expansion speed of the remnant at $t=t_{\rm nr}$.
Substituting Eq.~\ref{eq:e_vent_tnr} into Eq.~\ref{eq:rblisternr} one obtains
\begin{equation}
\label{eq:rblistermax}
\frac{R_{\rm bmax}}{R} \approx 0.028.
\end{equation}
This is an unexpected result and indicates that the maximum relative
size of a blister on the surface of a SNR does not depend on the
explosion energy, the ambient density or the ambient temperature. We
again see that the maximum size of a blister is comparable to the
shell thickness (since at $t=t_{\rm nr}$, $\delta R \approx R/33$).

\section{Analytical Results}
\label{sec:results}

In this section we use the analytical model developed in the previous
section to examine the key characteristics of blisters on the surfaces
of WBBs and SNRs. To allow a like-for-like comparison we assume that
the WBBs and SNRs expand into an ambient medium with $\rho_{0} =
10^{-24}\,{\rm g\, cm^{-3}}$ and $T_{0} = 10^{4}\,$K unless otherwise
noted. We assume that the mass fractions of hydrogen, helium and
metals are $X=0.705$, $Y=0.275$ and $Z=0.020$, respectively, which
gives an average particle mass of $0.615\,{\rm m_{H}}$ when the gas is
fully ionized. The corresponding sound speed for these conditions is
$c_{0}=15\,\kmps$.

\subsection{WBB Results}
\label{sec:wbb_results}

We consider first the evolution of blowouts around a bubble with the
following key parameters: $\dot{M} = 10^{-6}\,\Msolpyr$, $v_{\rm
  wind} = 2000\,\kmps$, $\rho_{\rm 0} = 10^{-24}\,{\rm g\,cm^{-3}}$,
$T_{\rm 0} = 10^{4}$\,K and $c_{\rm 0} = 15\,\kmps$. The rate of
energy injection, $\dot{E} = 1.26\times10^{36}\,\ergps$. The wind
injection parameters are typical of a single massive star. We refer to
this model as WBB-A.

\begin{table*}
\begin{center}
\caption[]{Assumed parameters for the WBB models and results. In all
cases we assume $v_{\rm wind} = 2000\,\kmps$.}
\label{tab:models}
\begin{tabular}{lllllllll}
\hline
\hline
Model & $\dot{E}$ & $\rho_{\rm 0}$ & $T_{0}$ & $c_{\rm 0}$ & $t_{\rm c}$ &
$t_{\rm nr}$ & $R_{\rm bmax}$ & $R_{\rm bmax}/R$ \\
 & (\ergps) & (${\rm g\,cm^{-3}}$) & (K) & (${\rm km\,s^{-1}}$) & (yr) &
 (yr) & (pc) & \\
\hline
WBB-A & $1.26\times 10^{36}$ & $10^{-24}$ & $10^{4}$ & 15 & 6600 & $7.76\times10^{5}$ & 1.22 & 0.17 \\ 
WBB-B & $1.26\times 10^{36}$ & $10^{-24}$ & $10^{2}$ & 1.5 & 11200 & $6.19\times10^{7}$ & 255 & 0.66 \\
WBB-C & $1.26\times 10^{36}$ & $10^{-24}$ & $10^{7}$ & 475 & 6600 & 13.8 & - & - \\
WBB-D & $1.26\times 10^{36}$ & $10^{-23}$ & $10^{4}$ & 15 & 1290 & $2.45\times10^{4}$& 0.38 & 0.17 \\
WBB-E & $1.26\times 10^{37}$ & $10^{-24}$ & $10^{4}$ & 15 & 12900 & $2.45\times10^{5}$ & 3.84 & 0.17 \\
\hline
\end{tabular}
\end{center}
\end{table*}

\begin{figure}
\psfig{figure=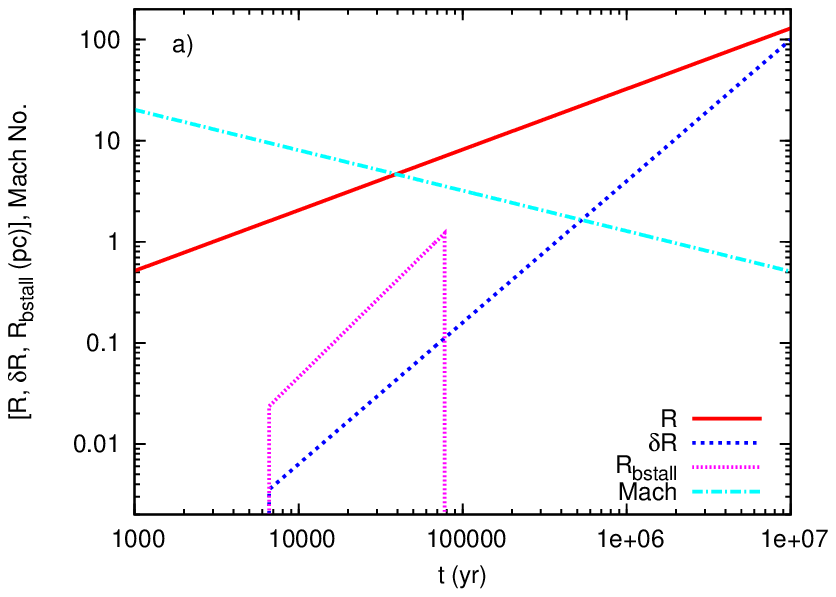,width=8.4cm}
\psfig{figure=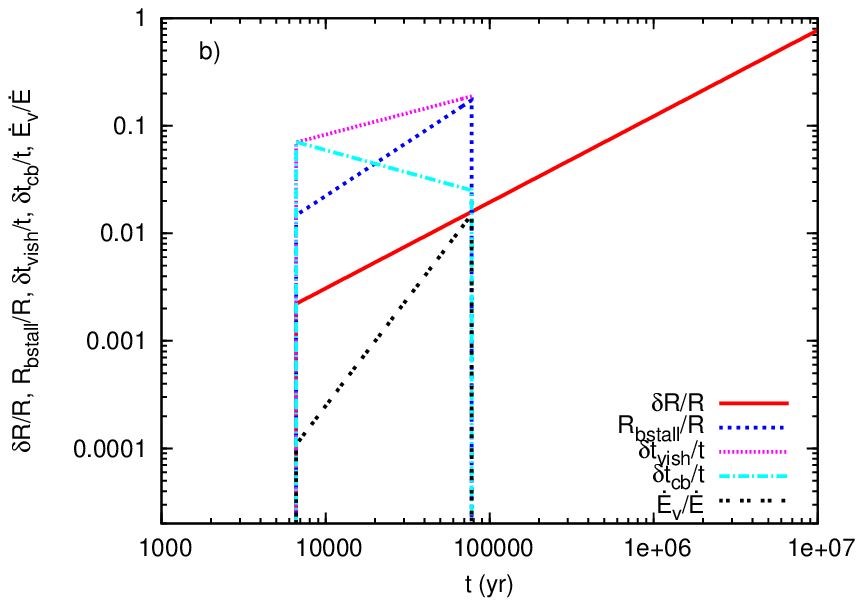,width=8.4cm}
\caption[]{Top: The WBB radius ($R$), Mach No. and shell thickness
  ($\delta R$), and the blister radius at the point its expansion
  stalls ($R_{\rm bstall}$), as a function of time for model
  WBB-A. Bottom: $\delta R/R$, $R_{\rm bstall}/R$, $\delta t_{\rm
    vish}/t$, $\delta t_{\rm cb}/t$ and $\dot{E}_{\rm v}/\dot{E}$ as a
  function of time for model WBB-A.}
\label{fig:modelA}
\end{figure}

Fig.~\ref{fig:modelA} shows that the bubble radius in model~WBB-A
grows to over 100\,pc by $10^{7}$\,yr. The thickness of the cold
swept-up shell grows from $\approx 3\times 10^{-3}$\,pc at the point
of its formation when the bubble age is 6565\,yr, to 38\,pc at
$5\times 10^{6}$\,yr. The increasing thickness of the cold shell is
caused by the continual decline in its Mach number as more and more
material is swept-up. The shell has a Mach number of 10 when the
bubble is 5760\,yr old, and a Mach number of 3.5 when it is 80,000\,yr
old. The shell becomes subsonic when the bubble is 1.82\,Myr
old. Vishniac instabilities grow very rapidly as soon as the shell is
formed ($\delta t_{\rm vish}/t \approx 0.07$), and so the shell is
likely to rupture soon afterwards. Blowouts will then occur and will
quickly develop and form cold shells ($\delta t_{\rm cb}/t \approx
0.07$). Hence the main WBB shell is indeed ``self-sealing''. Since
$\delta t_{\rm cb}/t$ declines as the age of the WBB increases this is
true at all times.

Blisters created soon after cold shell formation of the main WBB are
relatively small, however, because the rupture in the shell is also
small. Thus the earliest blisters stall once they have reached a
radius of $0.02$\,pc (this compares to a bubble radius of 1.6\,pc at
this time).  However, blisters have the opportunity to become
significantly larger as the shell slows and thickens, due to a
corresponding increase in size of the ruptures and the ``nozzle''
which inflates them. Fig.~\ref{fig:modelA} shows that blisters can
reach a maximum radius of 1.22\,pc at $t = 0.078$\,Myr. After this
time, the Vishniac instability can no longer rupture the shell. Thus
for model~WBB-A, the maximum blowout radius $R_{\rm bmax} = 0.173\,R$
(cf.  Eq.~\ref{eq:rblistermaxrel2}).  Each leakage event has an energy
flux of at most $0.015\,\dot{E}$.

The temperature of the hot gas within the main WBB is constant in time
(at least while this gas remains adiabatic), so the flow speed of gas
through ruptures is also independent of time. In model WBB-A, the
temperature of the interior is $\approx 4.5\times10^{7}$\,K, and gas
venting through a shell rupture has a velocity of $\approx
1000\kmps$. The venting gas (which is transonic with respect to its
own sound speed) has an adiabatic Mach number of 67 with respect to
the surrounding medium.  As we shall see, the gas venting from WBB
ruptures is expelled at much higher speeds and temperatures than gas
venting from SNR ruptures (cf. Sec.~\ref{sec:snr_results}).
 
\begin{figure}
\psfig{figure=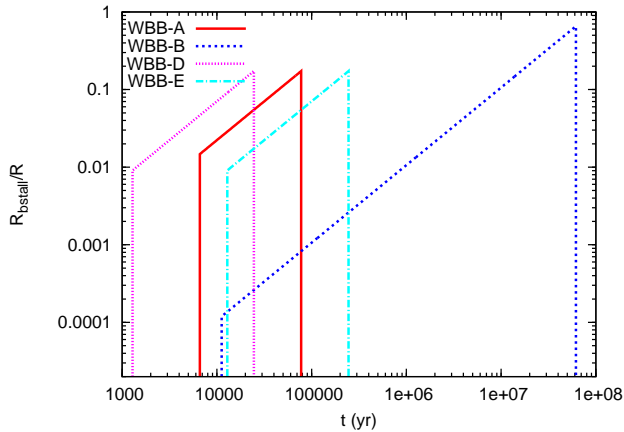,width=8.4cm}
\caption[]{The ratio of the radius of the stalled blister to the
  radius of the main WBB, $R_{\rm bstall}/R$, as a function of time
  for some of the models noted in Table~\ref{tab:models}.}
\label{fig:models}
\end{figure}

We have also investigated how the blisters depend on some key model
parameters. Table~\ref{tab:models} summarizes the different models
that were analyzed. We have varied the energy injection rate,
$\dot{E}$, and the ambient density and sound speed, $\rho_{\rm 0}$ and
$c_{\rm 0}$, respectively. The evolution of $R_{\rm bstall}/R$ for
these models is shown in Fig.~\ref{fig:models}. Reducing the ambient
sound speed (model~WBB-B) results in a higher Mach number for the
shell and consequently greater compression of the swept-up gas. This
allows the shell to stay thinner for longer, extending the timescale
over which the Vishniac instability may operate (up to a bubble age of
$t_{\rm nr}$). This in turn increases the maximum value of $R_{\rm
  bstall}/R$ to 0.66, as confirmed using
Eq.~\ref{eq:rblistermaxrel2}. However, to achieve this requires
continuous energy input for nearly 62 million years. If the lifetime
of the energy source is less than this (e.g., if the energy source is
a single massive star), the value of $R_{\rm bmax}/R$ will be smaller.

In contrast, setting $T_{\rm 0} = 10^{7}\,$K (model~WBB-C) prevents
any blowouts whatsoever. This is because the cold shell is already
subsonic at its formation (at 6600\,yr), and does not achieve the
necessary compression for the Vishniac instability to
operate\footnote{Note that it is likely that we slightly underestimate
  the cooling time for cold shell formation since we assume a strong
  shock.}. This is also manifest by the fact that $t_{\rm nr} < t_{\rm
  c}$.

Increasing the ambient density (model~WBB-D) causes the bubble
evolution to speed up, but does not alter the maximum value of $R_{\rm
  bstall}/R$. Indeed, Eq.~\ref{eq:rblistermaxrel2} shows that there is
no dependence on $\rho_{\rm 0}$. Increasing $\dot{E}$ (model~WBB-E)
causes an increase in the bubble size, speed and Mach number at any
specified time. However, if $\dot{M}$ is increased in proportion
(i.e. so that a specific value of $v_{\rm wind}$ is maintained), we
again see that the maximum value of $R_{\rm bstall}/R$ remains
unaffected. Only by further reducing $c_{\rm 0}$ or by increasing
$v_{\rm wind}$ are we able to increase the maximum value of $R_{\rm
  bstall}/R$. In reality, it may be difficult for a bubble to expand
into a neutral or molecular medium if the central star has a high
ionizing flux. One solution would be if the ionization front becomes
trapped by the shell \citep[e.g.,][]{Comeron:1997}. In such a scenario
the shell structure will be modified, and will consist of an inner
ionized part and an outer neutral part. The Vishniac instability may
then operate only in the outer neutral part of the shell. This is
clearly beyond the scope of the present work.

\begin{figure*}
\psfig{figure=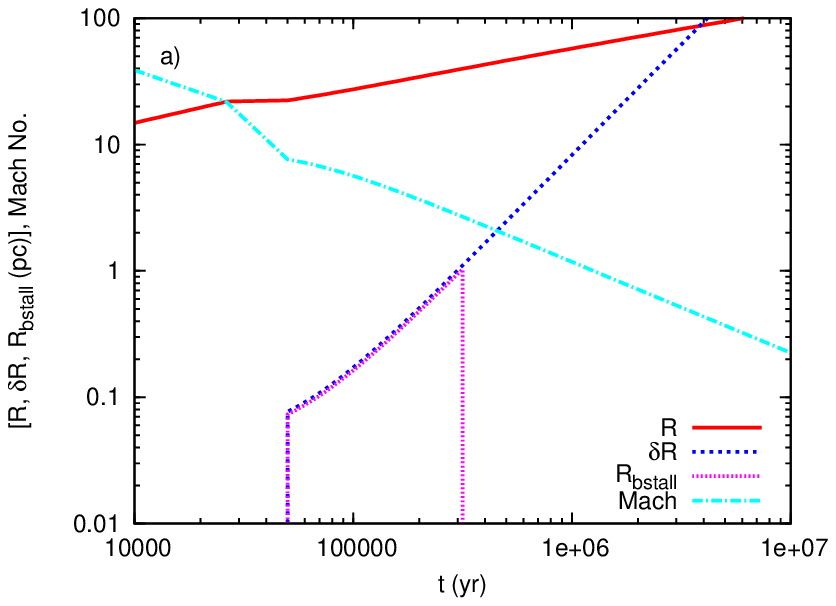,width=8.4cm}
\psfig{figure=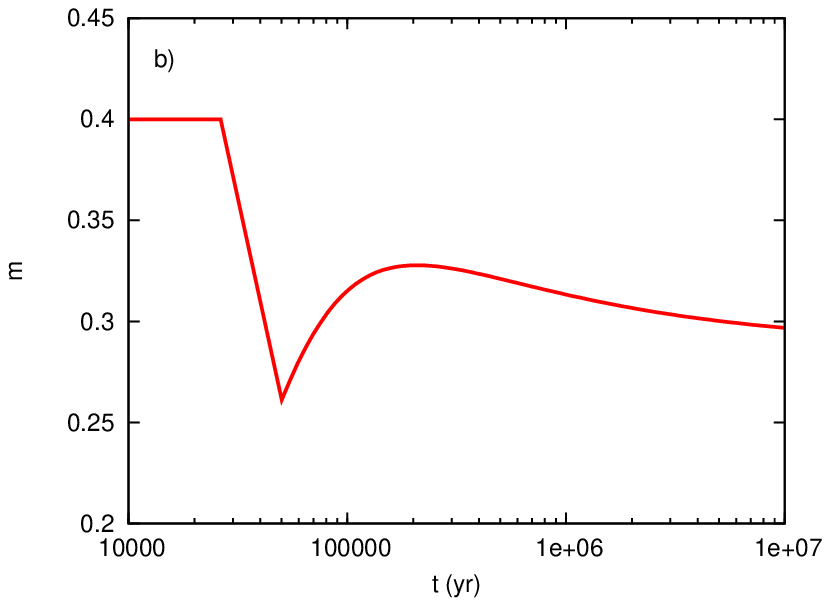,width=8.4cm}
\psfig{figure=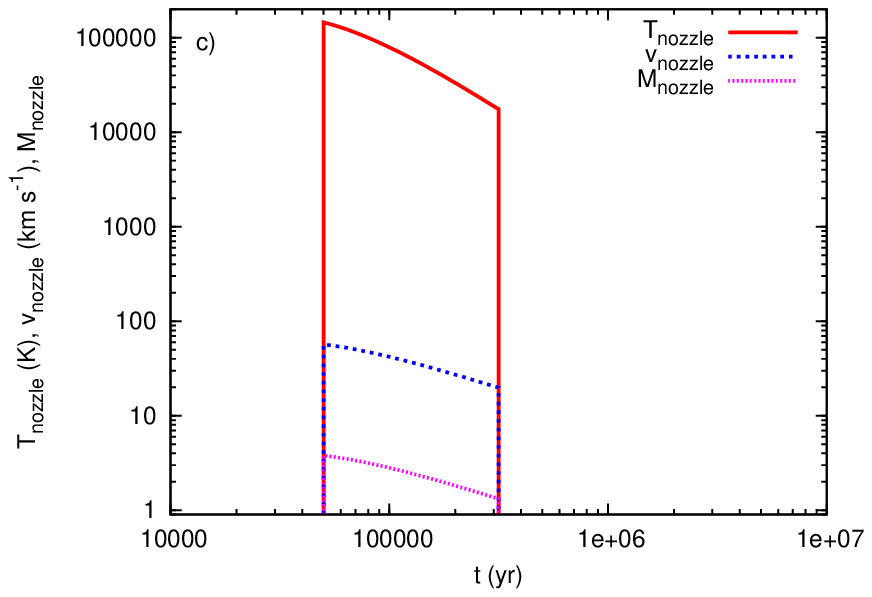,width=8.4cm}
\psfig{figure=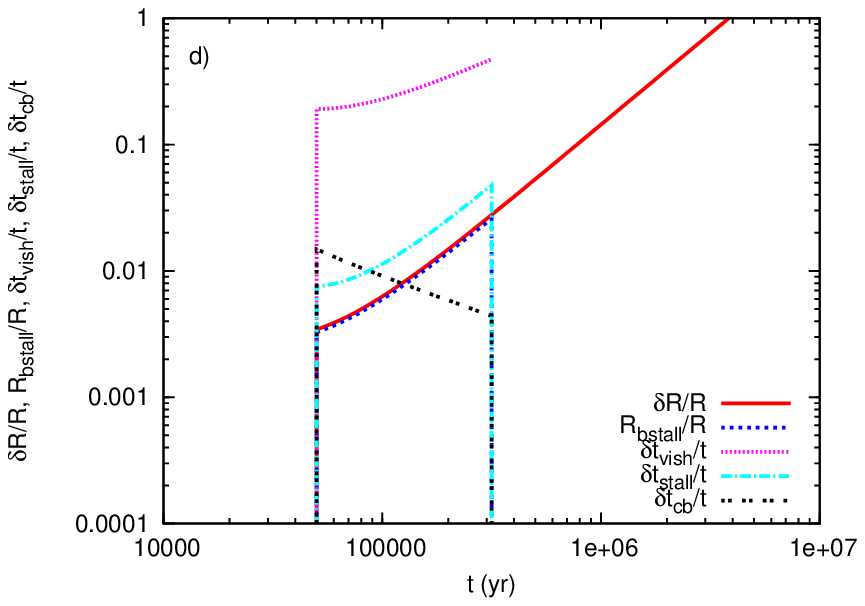,width=8.4cm}
\psfig{figure=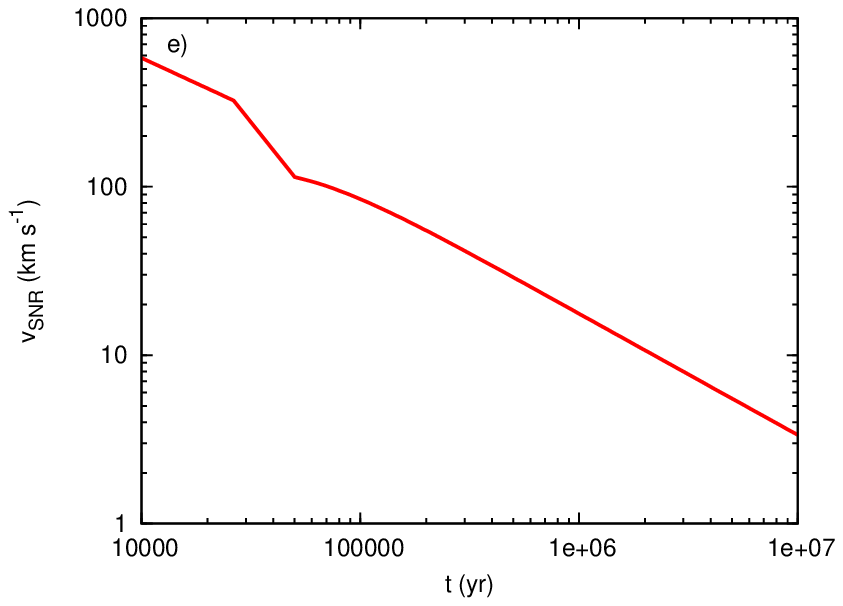,width=8.4cm}
\psfig{figure=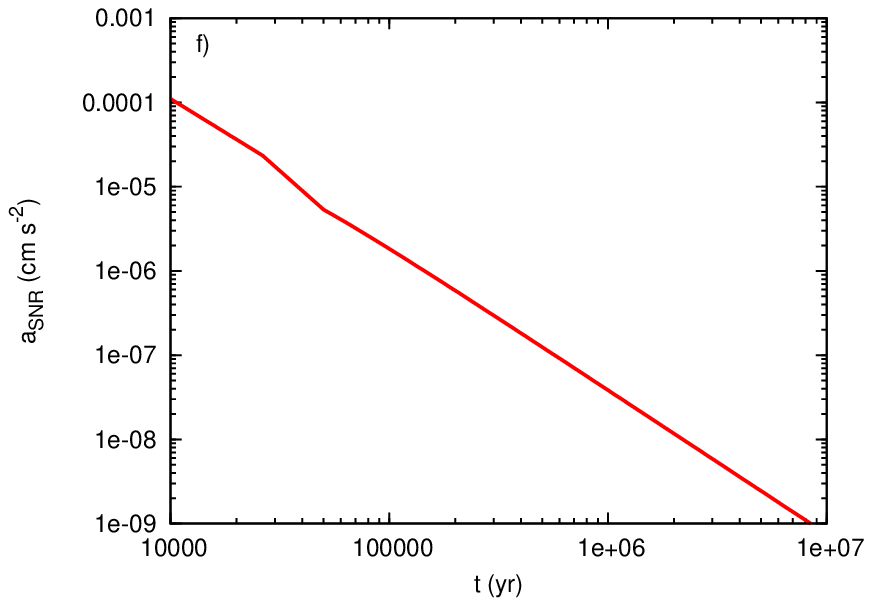,width=8.4cm}
\caption[]{a) The SNR radius ($R$), shell thickness ($\delta R$) and
  maximum blowout radius ($R_{\rm bstall}$) as a function of time for
  model SNR-A. The adiabatic Mach number of the shell is also
  shown. b) The deceleration index, $m$, of the remnant vs. time. c)
  The temperature, velocity and Mach number (with respect to the
  ambient medium) of the flow through a rupture vs. time (the gas flow
  through the rupture is transonic). d) $\delta R/R$, $R_{\rm
    bstall}/R$, $\delta t_{\rm vish}/t$, $\delta t_{\rm stall}/t$ and
  $\delta t_{\rm cool,b}/t$ as a function of time. e) The shell
  velocity vs. time. f) The shell deceleration rate vs. time.}
\label{fig:snrmodels}
\end{figure*}

\subsection{SNR Results}
\label{sec:snr_results}
We now consider the evolution of a SNR and the blowouts which may
develop. We first adopt the following key input parameters:
$E_{51}=1$, $\rho_{\rm 0} = 10^{-24}\,{\rm g\,cm^{-3}}$, $T_{\rm 0} =
10^{4}$\,K. With these parameters, $c_{\rm 0} = 15\kmps$, $\mu_{0} =
0.615$, and $n_{\rm H0} = 0.97\,\pcm3$. The remnant expands at its ST
rate until cooling becomes important. We refer to this as model
SNR-A - some key parameters are noted in Table~\ref{tab:snrmodels}.

Fig.~\ref{fig:snrmodels}a) shows that the remnant has formed a cool
shell by an age of $50,000$\,yr (cf.  Eq.~\ref{eq:t_trans}). Just
after this time the remnant has a radius of 22.4\,pc, an expansion
speed of $114\,\kmps$ and an adiabatic Mach number of 7.6. Note that
the expansion speed and Mach number drop significantly during the
process of cold shell formation as the remnant expansion stalls. The
deceleration parameter, $m$, drops from 0.4 (just prior to cold shell
formation the remnant expands according to the ST solution) to
$\approx0.26$ (Fig.~\ref{fig:snrmodels}b). The cold shell is initially
compressed quite thin ($\delta R = 0.077$\,pc) and is subject to the
Vishniac instability ($\delta t_{\rm vish} = 9590$\,yr).

Fig.~\ref{fig:snrmodels}c) shows that gas venting through a shell
rupture at the time of cold shell formation has a temperature $\approx
1.4\times 10^{5}$\,K and a velocity of 57\,\kmps. The venting gas has
an adiabatic Mach number of 3.8 with respect to the surrounding
medium. Hence it drives a shock into the undisturbed ISM and a blister
rapidly develops. Such blisters reach their maximum size $\approx
380$\,yr after the shell rupture and take 750\,yrs to form their own
cold shells (and thus ``self-seal'' the remnant) - see the plots of
$\delta t_{\rm stall}/t$ and $t_{\rm cb}/t$ in
Fig.~\ref{fig:snrmodels}d), respectively.  Note that blisters formed
from shell ruptures at later times form cold shells prior to
stalling. We note again that gas venting from WBB ruptures is expelled
at much higher speeds and temperatures than gas venting from SNR
ruptures (cf. Sec.~\ref{sec:wbb_results}). Note also that though
$\dot{R} > v_{\rm nozzle}$ for model SNR-A
(cf. Figs.~\ref{fig:snrmodels}c and~e), venting still occurs because
the gas just interior to the shell is itself expanding at a speed very
nearly equal to that of the shell.

Since $\delta t_{\rm vish}/t$ increases with time, the Vishniac
instability is more likely to rupture the shell soon after formation
of the cold shell.  However, if ruptures continue to occur blisters
grow to larger sizes before stalling as the remnant ages. At a time
near the post-cold-shell peak in the deceleration parameter (which
occurs at $t\sim0.2$\,Myr), we measure from Fig.~\ref{fig:snrmodels}a)
that $R_{\rm bstall}\propto t^{1.64}$ (Eq.~\ref{eq:rblisterattstall}
gives $R_{\rm bstall} \propto t^{1.67}$ for $m=0.33$). This difference
can be explained as due to the combination of the time dependence of
$m$ and our approximation of $P_{\rm s}$ in
Eq.~\ref{eq:p_postshock_highM}. The curvature in $R_{\rm bstall}/R$
reflects the time dependence of $m$ (cf. Eq.~\ref{eq:relsize_vs_t}).

Fig.~\ref{fig:snrmodels}a) shows that blisters can reach a maximum
radius of 1.03\,pc by $t=t_{\rm nr}=0.316$\,Myr, after which time the
Vishniac instability can no longer rupture the shell (which has a
radius $R = 39.8$\,pc). Thus for model SNR-A, the maximum blowout radius
is $\approx 0.026$\,R (cf. Eq.~\ref{eq:rblistermax}).
Each leakage event involves an energy flux of at most
$5.9\times10^{33}\ergps$.

\begin{figure}
\psfig{figure=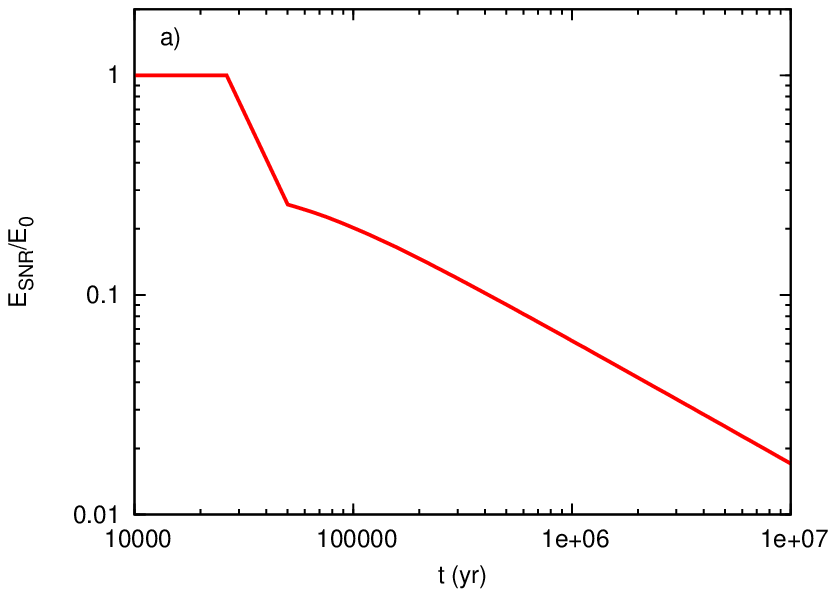,width=8.4cm}
\psfig{figure=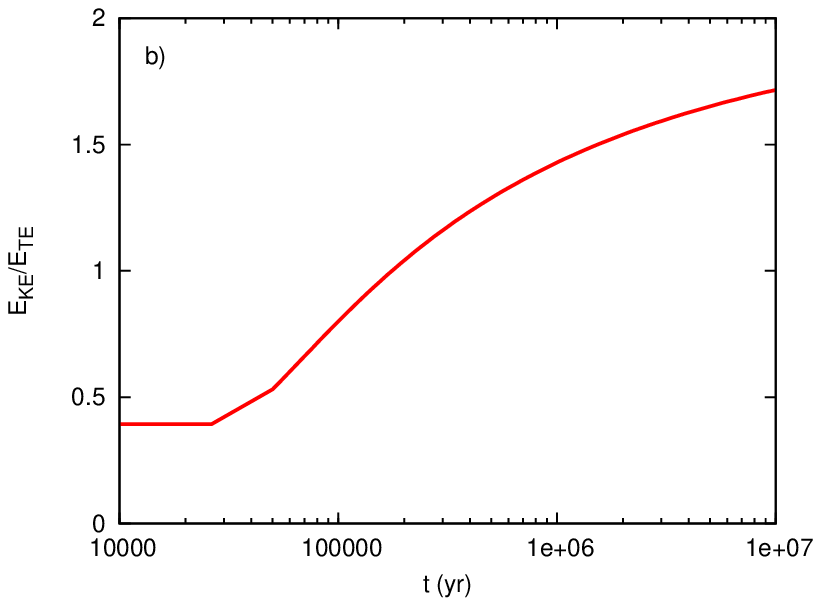,width=8.4cm}
\caption[]{a) The energy of the remnant vs. time for model SNR-A. b) The ratio
  of the kinetic to thermal energy of the remnant vs. time.}
\label{fig:snrenergy}
\end{figure}

\citet{Bandiera:2004} also provide equations for the ratio of the
kinetic to thermal energy within the remnant (their Eq.~16) and for
the total energy of the remnant (their Eq.~19). Note that the
equations do not account for the thermal energy swept up by the
remnant, which can be significant at late times.  In
Fig.~\ref{fig:snrenergy}a) we plot the ratio of the current energy of
the remnant compared to its initial energy. The remnant radiates away
75 per cent of its energy during the process of cold shell formation
\citep[in good agreement with numerical simulations - see,
e.g.,][]{Falle:1981,Pittard:2003} and contains just 11 per cent of its
energy at $t=t_{\rm nr}$, 6 per cent at $t=1$\,Myr, and only 5 per
cent when the shell expansion becomes subsonic at
$t=1.25$\,Myr. During this interval the kinetic energy fraction of the
remnant increases from 28 per cent to 60 per cent of the total remnant
energy (Fig.~\ref{fig:snrenergy}b).

\begin{table*}
\begin{center}
{\small
  \caption[]{Assumed parameters for the SNR models and
    results. Subscript $tr$ indicates when the remnant transitions
    from adiabatic to radiative. A cool shell is assumed to form at
    $t=1.6\,t_{\rm tr}$ (see Sec.~\ref{sec:analytical_snrs} for
    details). Subscript $nr$ indicates when the cold shell becomes too
    thick for the Vishniac instability to continue to operate. $M_{\rm
      su}$ is the swept-up mass at $t=t_{\rm tr}$. Model SNR-G forms a
    cold shell at a Mach number which is too low for the Vishniac
    instability to operate.}
\label{tab:snrmodels}
\begin{tabular}{llllllllllllll}
\hline
\hline
Model & $E_{51}$ & $\rho_{\rm 0}$ & $T_{0}$ & $c_{\rm 0}$ & $t_{\rm tr}$ & $R_{\rm tr}$ & $v_{\rm tr}$ & Mach$_{\rm tr}$ & $M_{\rm su}$ & $t_{\rm  nr}$ & $R_{\rm nr}$ & $R_{\rm bmax}$ & $R_{\rm bmax}/R$ \\
 & ($10^{51}$\,ergs) & (${\rm g\,cm^{-3}}$) & (K) & (${\rm km\,s^{-1}}$) & (yr) &
 (pc) & (${\rm km\,s^{-1}}$) & & ($\Msol$) &
 (yr) & (pc) & (pc) & \\
\hline
 SNR-A & 1 & $10^{-24}$   & $10^{4}$ & 15 & 29400 & 19.3 & 259 & 17.3 & 1006 & $3.16\times10^{5}$ & 39.8 & 1.03 & 0.0259\\ 
 SNR-B & 1 & $10^{-23}$   & $10^{4}$ & 15 & 8700  & 7.49  & 340 & 22.7 & 586  & $1.45\times10^{5}$ & 17.8 & 0.50 & 0.0280\\
 SNR-C & 1 & $10^{-22}$   & $10^{4}$ & 15 & 2570  & 2.90  & 445 & 29.8 & 341  & $6.22\times10^{4}$ & 7.76 & 0.22 & 0.0278\\
 SNR-D & 10 & $10^{-24}$  & $10^{4}$ & 15 & 50600 & 38.0 & 297 & 19.8 & 7680 & $7.00\times10^{4}$ & 85.1 & 2.40 & 0.0282\\
 SNR-E & 0.1 & $10^{-24}$ & $10^{4}$ & 15 & 17100 & 9.82  & 226 & 15.1 & 132  & $1.53\times10^{5}$ & 19.1 & 0.50 & 0.0263\\
 SNR-F & 1 & $10^{-24}$   & $10^{2}$ & 1.5 & 43600 & 26.2 & 238 & 230  & 1200 & $1.78\times10^{7}$ & 166  & 4.52 & 0.0273\\
 SNR-G & 1 & $10^{-24}$   & $10^{7}$ & 475 & 29400 & 19.3 & 259 & 0.55 & 1007 & $-$ & $-$ & $-$ & $-$ \\
\hline
\end{tabular}
}
\end{center}
\end{table*}

\begin{figure}
\psfig{figure=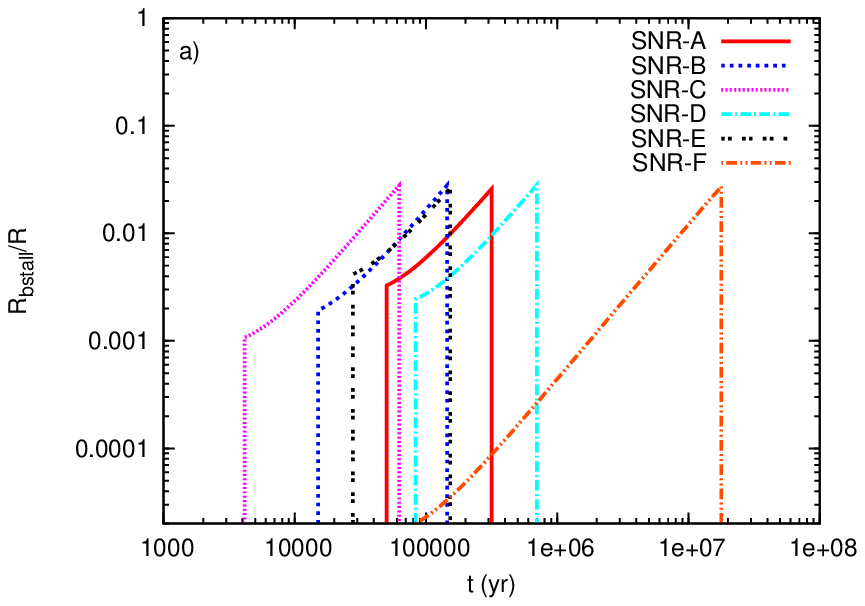,width=8.4cm}
\psfig{figure=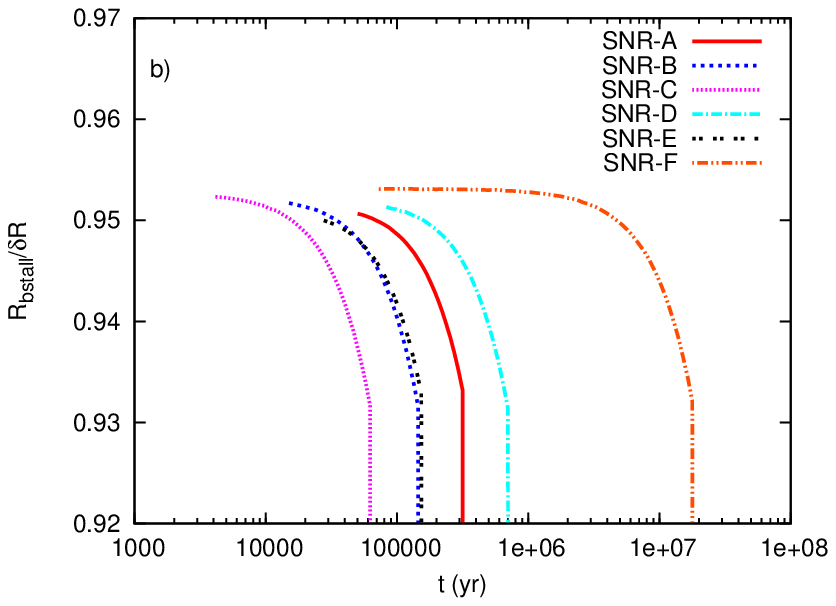,width=8.4cm}
\caption[]{a) The ratio of the blister stall radius to the SNR radius ($R_{\rm
    bstall}/R$) as a function of time for some of the models noted
  in Table~\ref{tab:snrmodels}. b) The ratio of the blister stall radius to the
  shell thickness ($R_{\rm bstall}/\delta R$) as a function of time.}
\label{fig:snrmodels2}
\end{figure}

We have also investigated how the SNR blisters depend on some key
model parameters. Table~\ref{tab:snrmodels} summarizes the different
models which were analyzed. We have varied the initial mechanical
explosion energy, $E_{\rm 51}$, and the ambient density and sound
speed, $\rho_{\rm 0}$ and $c_{\rm 0}$, respectively. The evolution of
$R_{\rm bstall}/R$ for these models is shown in
Fig.~\ref{fig:snrmodels2}a). Changing the values of $E_{51}$ and
$\rho_{\rm 0}$ (compare models SNR-A to SNR-E) varies the rapidity of
the remnant evolution and the Mach number of the shell at the time of
cold shell formation. Therefore the stall radii of the blisters at the
time of cold shell formation is different between these
models. However, each model has the same maximum blowout radius
relative to the remnant radius at $t=t_{\rm nr}$ ($R_{\rm bmax}/R$),
which is expected given that Eq.~\ref{eq:rblistermax} has no
dependence on $E_{\rm 51}$, $\rho_{\rm 0}$ or $c_{\rm 0}$.

Remnants which expand into a cold medium are able to maintain
supersonic shells for a considerably longer time interval (see
model~SNR-F). In contrast, remnants which expand into a very hot
medium have shells with reduced Mach numbers. This can result in the
Mach number of the cold shell at its formation being too low for the
Vishniac instability to grow, thus preventing blowouts and blisters
(model~SNR-G).
 
Fig.~\ref{fig:snrmodels2}b) shows the value of $R_{\rm bstall}/\delta
R$ for our models. For shells at high Mach number the models show that
$R_{\rm bstall}/\delta R\approx0.95$ (cf.
Eq.~\ref{eq:rblister_over_dR}). The value of this ratio decreases as
the Mach number of the shell drops and the thermal pressure becomes
important in the shock jump conditions.

\subsection{Comparison of WBB and SNR Results}
\label{sec:wbb_snr_comparison}

\begin{figure*}
\psfig{figure=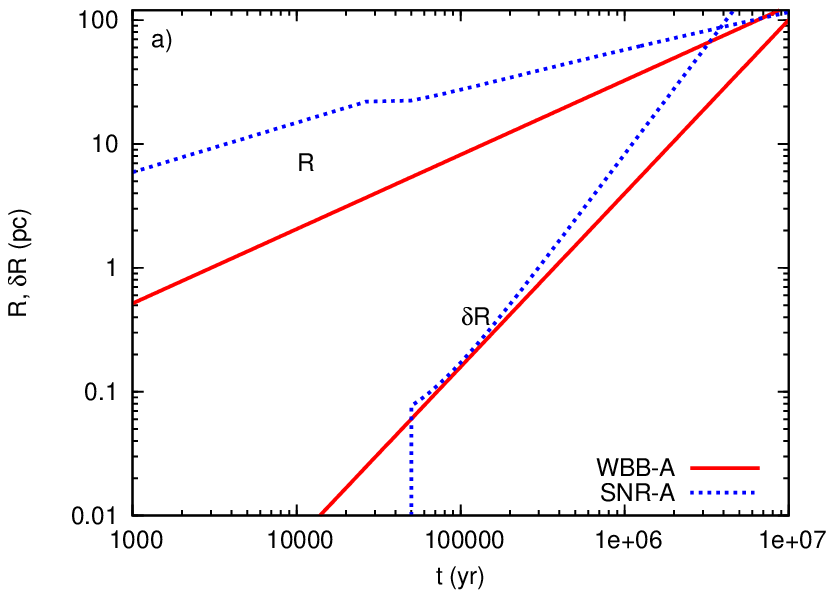,width=8.4cm}
\psfig{figure=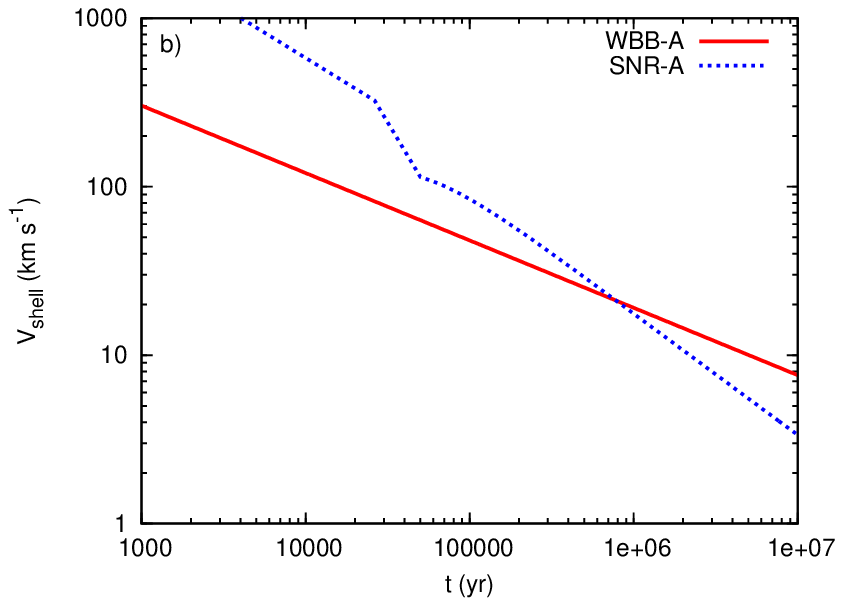,width=8.4cm}
\psfig{figure=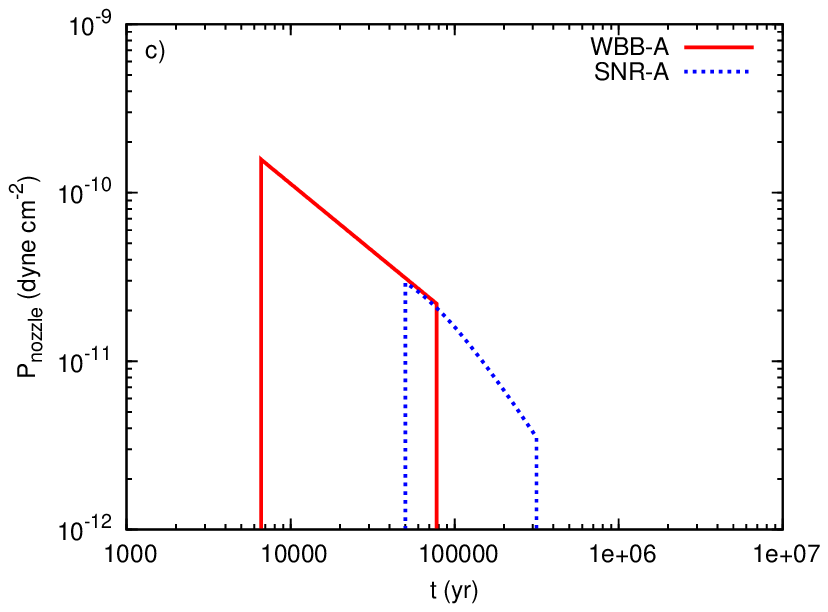,width=8.4cm}
\psfig{figure=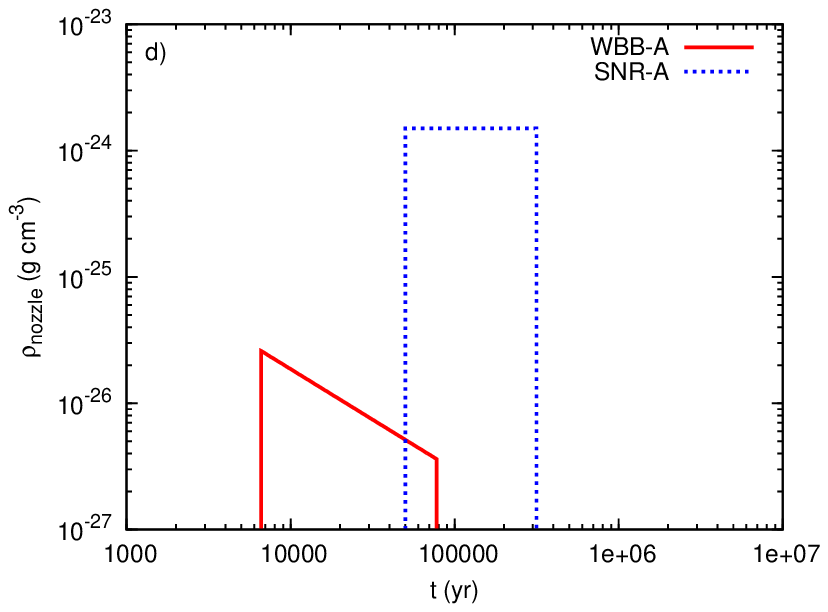,width=8.4cm}
\psfig{figure=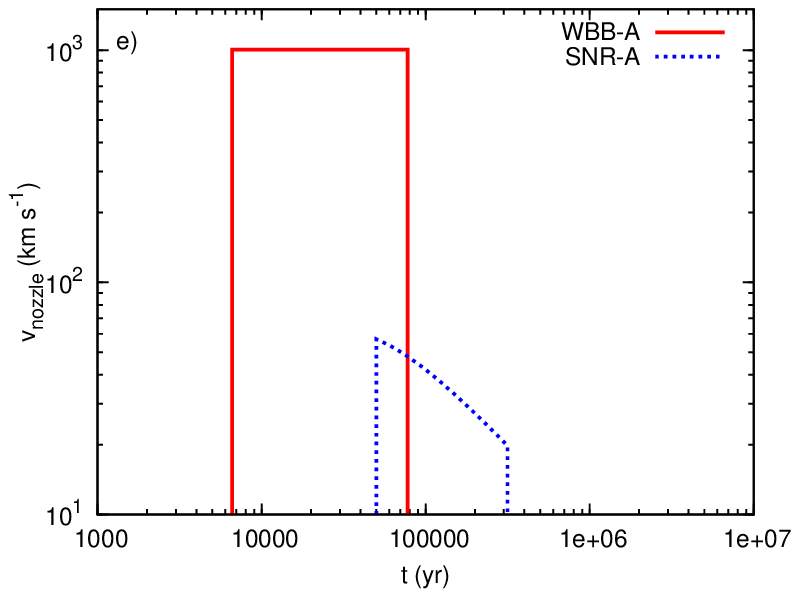,width=8.4cm}
\psfig{figure=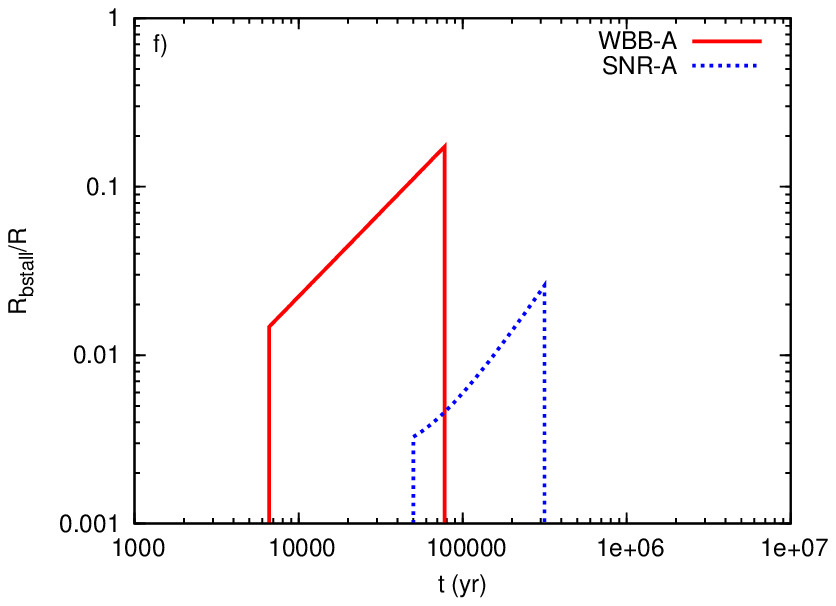,width=8.4cm}
\caption[]{Comparison of models WBB-A and SNR-A. Shown as a function
  of time are a) the radius of the shell and its thickness; b) the
  shell expansion velocity; c-e) the pressure, density and velocity of
  gas as it flows through a shell rupture; f) the size of blisters
  relative to the radius of the main shell.}
\label{fig:wbb_snr_comparison}
\end{figure*}

We have seen from the work in the previous subsections that the
relative size of blisters compared to their WBB or SNR host is greater
for WBBs. We wish to understand why this is
so. Fig.~\ref{fig:wbb_snr_comparison} compares various properties of
models WBB-A and SNR-A. These models have the same ambient density
$\rho_{0}$, temperature $T_{0}$, and sound speed $c_{\rm 0}$.

Fig.~\ref{fig:wbb_snr_comparison}a) and b) show that the SNR model has
a relatively larger shell radius during the period of interest when
shell ruptures can occur, and that the velocity of the shell is also
higher for the majority of this period too. We also see that the SNR
model has a thicker cold shell than the WBB model at any given time
after its formation (Fig.~\ref{fig:wbb_snr_comparison}a). The pressure
of the gas flowing through the shell rupture is roughly comparable
during the period that shell ruptures occur in both models (roughly
from $50,000-80,000$\,yr - see
Fig.~\ref{fig:wbb_snr_comparison}c). However, because shell ruptures
start when the SNR is older the typical pressure of the flow through a
rupture is less in the SNR model. More striking is the fact that the
density of gas flowing through the rupture is much lower, and it flows
out much more rapidly, in the WBB model
(Figs.~\ref{fig:wbb_snr_comparison}d and~e). Finally,
Fig.~\ref{fig:wbb_snr_comparison}f) shows the relative size of
blisters as a function of time for both models.

Eq.~\ref{eq:edotvent} gives the rate of energy venting through a shell
rupture. With our assumption that the flow through the
rupture is transonic ($v = c$), we have
\begin{equation}
\dot{E}_{\rm v} = \frac{7}{3}{P c A}
\end{equation}
when $\gamma=5/3$. With $A \approx \pi^{3}\delta R^{2}$ and $\delta R
= R/(5M^{2})$, we find using Eq.~\ref{eq:rblisterstall1.5} that
\begin{equation}
\label{eq:relsize1}
\frac{R_{\rm bstall}}{R} \approx 0.18\,\left(\frac{P
    c}{\rho_{0}}\right)^{1/2} c_{0}^{2} \dot{R}^{-7/2}.
\end{equation}
Since $c = \sqrt{\gamma P/\rho}$, this expression can also be written
as
\begin{equation}
\label{eq:relsize2}
\frac{R_{\rm bstall}}{R} \approx 0.21\,P^{3/4} \rho^{-1/4} \rho_{0}^{-1/2} c_{0}^{2} \dot{R}^{-7/2}.
\end{equation}
Although the dependence on $\rho$ is weakest out of all of these
variables, it is the one whose value is most different between the two
models. In the SNR model it is assumed to be constant with time and
equal to $1.5\,\rho_{0}$, whereas in the WBB model it is assumed to be
equal to the average density of the hot interior gas
(Eq.~\ref{eq:rhobub}).

Table~\ref{tab:wbbsnrcomp} notes values of interest from models WBB-A
and SNR-A at the time of cold shell formation ($t_{\rm c}$) and at the
time that the Vishniac instability ceases to operate ($t_{\rm
  nr}$). We see that the $\dot{R}^{-7/2}$ term in
Eqs.~\ref{eq:relsize1} and~\ref{eq:relsize2} favours (by about a
factor of 2) a larger value of $R_{\rm bstall}/R$ for the SNR
model. However, the $(Pc)^{1/2}_{\rm nozzle}$ term favours a larger
value of $R_{\rm bstall}/R$ for the WBB model by a factor of 10 or
more and so clearly dominates this comparison. 

Both $P$ and $c$ contribute to the greater value of $(Pc)^{1/2}_{\rm
  nozzle}$ from the WBB model. However, inspection of the values for
$P$ and $c$ in Table~\ref{tab:wbbsnrcomp} reveals that it is the
higher value of $c$ which dominates. Hence, the greater
relative size of blisters on WBBs is ultimately due to the much higher
temperature of the hot interior gas in the WBB. This leads to a much
faster flow through the rupture and consequently to a much higher energy
flux through unit area of a rupture into the developing
blister. Because the size of the rupture is related to the thickness
and thus also to the radius of the shell, we see that the greater
relative energy flux through a rupture in a WBB shell allows a larger
blister to be blown for a given shell radius.

\begin{table*}
\begin{center}
{\small
\caption[]{Comparison of the WBB and SNR parameters at given
  times. The ambient medium has identical properties in both models:
  $\rho_{0} = 10^{-24}\,{\rm g\, cm^{-3}}$, $T_{0} = 10^{4}$\,K and
  $c_{0} = 15\kmps$. For model WBB-A, $t_{\rm c} = 6603$\,yr and
  $t_{\rm nr} = 0.776$\,Myr. For model SNR-A, $t_{\rm c} = 47,100$\,yr
  ($t_{\rm tr} = 29,400$\,yr) and
  $t_{\rm nr} = 0.316$\,Myr. Note that $c_{\rm nozzle}$ is less than
  the shell expansion velocity, $\dot{R}$, at all times in model SNR-A.}
\label{tab:wbbsnrcomp}
\begin{tabular}{lllllllllll}
\hline
\hline
Model & t & R & $\dot{R}$ & $P_{\rm nozzle}$ & $\rho_{\rm nozzle}$ &
$T_{\rm nozzle}$ & $c_{\rm nozzle}$ & $M_{\rm nozzle}$ & $(P c)^{1/2}_{\rm nozzle}$ & $R_{\rm bstall}/R$ \\
      & & (pc)& ($\kmps$) & (${\rm dyn\, cm^{-2}}$) & (${\rm
        g\,cm^{-3}}$) & (K) & ($\kmps$) & & (${\rm g^{1/2} s^{-3/2}}$)
      & \\
\hline
WBB-A & $t_{\rm c}$ & 1.60 & 142 & $1.57\times10^{-10}$ &
$2.6\times10^{-26}$ & $4.52\times10^{7}$ & 1005 & 67 & $4.0\times10^{-4}$ & 0.0147 \\
      & $t_{\rm nr}$ & 7.03 & 53.1 & $2.19\times10^{-11}$ &
      $3.6\times10^{-27}$ & $4.52\times10^{7}$ & 1005 & 67 &
      $1.5\times10^{-4}$ & 0.173 \\
\hline
SNR-A & $t_{\rm c}$ & 22.4 & 114 & $2.91\times10^{-11}$ & $1.5\times10^{-24}$ &
$1.45\times10^{5}$ & 57 & 3.8 & $4.1\times10^{-5}$ & 0.0033 \\
      & $t_{\rm nr}$ & 39.8 & 40.2 & $3.53\times10^{-12}$ &
      $1.5\times10^{-24}$ & $1.75\times10^{4}$ & 19.8 & 1.3 &
      $8.4\times10^{-6}$ & 0.0259 \\
\hline
\end{tabular}
}
\end{center}
\end{table*}

%%%%%%%%%%%%%%%%%%%%%%%%%%%%%%%%%%%%%%%%%%%%%%%%%%%%%%%%%%%%%%%%%%%%%%%%%%%%%%%
\section{Discussion}
\label{sec:discussion}

\subsection{WBB Results}

\subsubsection{Comparison to Numerical Simulations}
\label{sec:numericalsims}
In recent years there have been many published works on hydrodynamical
simulations of WBBs. The Vishniac instability can be clearly
identified in numerical simulations of a wind blowing into a uniform
medium \citep[e.g.,][]{Strickland:1998,Dwarkadas:2001,Freyer:2006},
and in simulations which study the interaction of bubbles blown by
massive stars \citep[e.g.,][]{vanMarle:2012,Krause:2013}. In all of
these works the swept-up shell is decelerating and is therefore stable
to RT instabilities, which makes identifying the Vishniac instability
quite straightforward.  However, in situations where RT instabilites
also occur \citep[such as where a faster wind blows into a slower
wind,
e.g.,][]{Garcia-Segura:1995,Garcia-Segura:1996,vanVeelen:2009,Toala:2011},
identification of the Vishniac instability is less secure.

It is interesting that \citet{Dwarkadas:2007} notes that the Vishniac
instability seen in his simulations was stronger in earlier models
\citep[e.g.][]{Dwarkadas:2001}. Since the latter were at lower
resolution he expressed confusion at this finding. However, the
assumed temperature of the ambient medium was different in the
simulations ($\sim 100$\,K vs. $\sim 8000$\,K). We suggest that this
difference is likely to be the underlying cause, since a lower ambient
temperature keeps the Mach number of the shell higher, which results
in a longer timescale over which the growth of the instabilities can
operate (and ultimately a greater relative blister radius
cf. Eq.~\ref{eq:rblistermaxrel2}).

\begin{figure*}
\psfig{figure=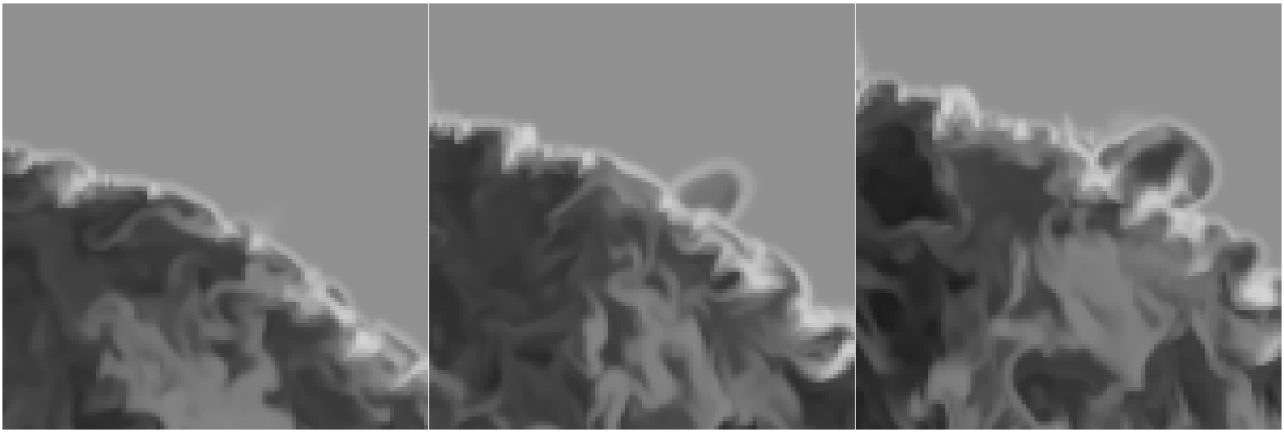,width=17.0cm}
\caption[]{Time series of blister growth from a 3D hydrodynamic
  simulation of a WBB. The blister can be seen growing from a rupture
  in the swept-up shell located near the centre of the left
  panel. The density scale is logarithmic between $10^{-27}\,{\rm g\,
    cm^{-3}}$ (black) to $10^{-19}\,{\rm g\, cm^{-3}}$ (white). The
  spatial scale of each panel is $8\times8$\,pc, and the size of each
  grid cell is 0.0625\,pc (thus there are $128\times128$ grid cells visible
  in each panel). The centre of the WBB
  is 9.75\,pc below the bottom left corner of each panel. The time of each
  snapshot (left to right) is 0.697\,Myr, 0.761\,Myr and
  0.823\,Myr. The simulation used the same setup as SimA from
  \citet{Rogers:2013} but with the ambient density increased by a
  factor of 100. The large density variations within the bubble are
  extremely time-dependent and result from the ablation of dense gas
  closer to the stellar cluster.}
\label{fig:sim}
\end{figure*}

Table~\ref{tab:literatureresults} summarizes work in the literature
where Vishniac instabilities are noted in hydrodynamic simulations of WBBs. The
main parameters adopted for the simulations are noted. We also note
the time for cold shell formation, $t_{\rm c} \approx (2.3 \times
10^{4}\,{\rm yr}) n_{\rm 0}^{-0.71}\,\dot{E}_{38}^{0.29}$, and the time when the
Vishniac instability ceases to operate, $t_{\rm nr}$ (Eq.~\ref{eq:t_novish}).
We can now study the published results in the light of these
values. 

\citet{Strickland:1998} note that cooling of the swept-up gas and the
process of shell formation takes several 1000 years, beginning at a
bubble age of about 4000 years. This is in good agreement with the
estimate of $t_{\rm c}$ in Table~\ref{tab:literatureresults}. While
the Vishniac instability may operate up to a bubble age of about
$0.2$\,Myr for the parameters adopted, the main focus of the
\citet{Strickland:1998} paper is not on the long-term evolution of the
WBB, and the latest time at which a 2D plot of the density is shown is
14630\,yr (see the bottom-right panel of their Fig.~1). The Vishniac
instability is clearly seen at this time. We estimate that the
relative scale of the shell perturbations, $R_{\rm p} = (R_{\rm
  max}-R_{\rm min})/\overline{R} \approx 0.04$, where $R_{\rm max}$
($R_{\rm min}$) is the maximum (minimum) radius of any part of the
cold shell and $\overline{R}$ is its average radius. The inferred
value for $R_{\rm p}$ is larger than the expected size of blowouts at
this time ($R_{\rm bstall}/R = 0.015$). We conclude, therefore, that
the perturbations are not blowouts and that their larger than expected
size results from under-resolving the shell: at $t=14630$\,yr we
expect a shell Mach number $M=9.6$, a bubble radius $R=3.47$\,pc, and
a shell thickness $\delta R = 0.0077$\,pc (hence $\delta R/R \approx
2\times 10^{-3}$), whereas the size of the grid cells is $0.0162$\,pc.

The Vishniac instability is also seen in the simulations of
\citet{Dwarkadas:2001}, but unfortunately the information provided in
this work is a little sparse. Nevertheless, using the values noted in
Table~\ref{tab:literatureresults} we estimate that shell formation
occurs at 4200\,yrs and that the Vishniac instability could
potentially operate until a bubble age of 20\,Myr, aided by the lower
sound speed of the ambient medium. Fig.~3 in \citet{Dwarkadas:2001}
clearly shows Vishniac instabilities and there seems also to be
evidence for shell ruptures and blisters. At $t=4.5$\,Myr we estimate
from his Fig.~3 that $R_{\rm p} \approx 0.09$, though we predict that
$R_{\rm bstall}/R = 0.18$ at this time.

\citet{Freyer:2006} note that the Vishniac instability ceases early in
the evolution of their simulated WBB (by $t=0.2$\,Myr), due to the
increase in the shell thickness (note that the WBB expands into a
medium with $T \approx 8000$\,K because of the presence of a HII
region around their bubble). We estimate that $t_{\rm c} = 810$\,yr
and $t_{\rm nr} \approx 0.025$\,Myr. Their Fig.~3 (at $t=700$\,yr)
shows the existence of a shell but it is clear that at this stage it
is too thick for rapid growth of the Vishniac instability, perhaps
because the shell is still cooling and has yet to be fully
compressed. However, the numerical resolution could also be a factor
(there are only 32 cells per bubble radius and the shell thickness is
about 5 cells).  In contrast, their Fig.~4 (at $t=0.05$\,Myr) is
already after the window of opportunity for the Vishniac
instability. Despite these issues small perturbations are visible in
both of their Figs.~3 and~4.

\citet{Ntormousi:2011} investigate the interaction of the winds from
several massive star clusters. Each cluster contains 50 stars with
time-dependent mass and energy injection rates. During the first
3\,Myr we estimate that each cluster has an average $\Mdot \approx
3\times10^{-5}\,\Msolpyr$ and $\dot{E} \approx
7.7\times10^{37}\,\ergps$ (and therefore an average $v_{\rm wind} \approx
2850\,\kmps$). We determine that the Vishniac instability
should occur up to a cluster age of about $0.79$\,Myr. Unfortunately
the earliest plot of the density is at 3\,Myr but the shell is clearly
very unstable even at this time (see their Fig.~2). The authors
note that 3 instabilities are simultaneously at work: Vishniac,
thermal and Kelvin-Helmholtz. However, since the shocked gas can cool
to a lower temperature than the pre-shock gas we believe that the behaviour
of the shell is dominated by the thermal instability.

By far the best example of the Vishniac instability in numerical
simulations of WBBs in the literature is in
\citet{vanMarle:2012}. Fig.~1 in this paper shows the main-sequence
WBB around a $40\,\Msol$ star at $t=4.3$\,Myr. Vishniac instabilites
are clearly seen (and expected) at this time. It is more difficult,
even with the aid of the online movies, to decide whether the shell
also ruptures and blisters form, but it is possible that this occurs
in several places around the bubble. In any case we measure $R_{\rm p}
\approx 0.18$, in reasonable agreement with the predicted $R_{\rm
  bstall}/R = 0.30$ at this time. The numerical resolution is good,
with about 250 cells per bubble radius at $t=4.3$\,Myr.

\citet{Krause:2013} investigate the collision of 3 WBBs.  Their
simulations are 3D and thus capture important new physics, but this
means that they also suffer from poor resolution compared to some of
the other 2D simulations described above. We estimate that up until
$t=1.95$\,Myr, the 60\,\Msol~star has an average mass-loss rate of
$\Mdot \approx 3\times10^{-6}\,\Msolpyr$ and injects into the ISM an
average energy flux of $\dot{E} \approx 10^{37}\,\ergps$ (hence an
average $v_{\rm wind} \approx 3240\,\kmps$). Using these values as
input we find that the Vishniac instability could in theory occur up
until a bubble age of $\approx30$\,Myr (though a $60\,\Msol$ star will
not live this long). The shell perturbations seen around the WBB of
the 60\,\Msol~star shown in the top right panel of their Fig.~3 at
$t=1.95$\,Myr have $R_{\rm p} \approx 0.25$. This is nearly 5 times
greater than the predicted value of $R_{\rm bstall}/R \approx
0.053$. This discrepancy, like that of \citet{Strickland:1998}, may
result from the low resolution (there are roughly 60 cells per bubble
radius at this time, whereas we expect $\delta R/R \approx 2\times
10^{-3}$).

Finally we note that we have seen the Vishniac instability in our own
simulations of WBBs and what we believe is the first unambiguous
evidence for shell rupture and blowouts/blister formation. This
behaviour was first reported by \citet{Rogers:2013} and is shown here
in Fig.~\ref{fig:sim}.  Although the (cluster) wind initially blows
into a highly inhomogeneous medium, a roughly spherical bubble forms
after this blows out into the surrounding uniform medium\footnote{At
  these later stages the flow has some similarities to a standard WBB
  (e.g., there is a reverse shock which heats the wind to high
  temperatures while a forward shock sweeps up and compresses the
  ambient medium into a thin shell). However, the centre of the bubble
  contains a large reservoir of cold, dense gas which is slowly
  ablated by the cluster wind and mixed into the bubble interior, and
  which significantly affects the bubble structure, morphology and
  evolution \citep[see also][]{Pittard:2001a,Pittard:2001b}.}.
Fig.~\ref{fig:sim} focuses on a small part of the swept-up shell when
the cluster wind has been blowing for $0.7-0.8$\,Myr.  The left-most
panel shows the initial rupture of the swept-up shell (the shell has a
width of $\delta R = 2-3$ grid cells). The hot gas in the bubble
interior then vents through this opening, blowing the blister which is
seen expanding in the other two panels. At later times the identity of
the blister is compromised by the growth of other nearby
blisters. Unfortunately the nature of this simulation means that it is
not possible to run the analysis in Sec.~\ref{sec:analytical_model} on
it. Note also that when the ambient temperature is $8000$\,K, the
swept-up shell remains too thick for the Vishniac instability to
efficiently operate and no blisters occur \citep[cf. Fig.~1
in][]{Rogers:2013}.

\begin{table*}
\centering
\caption{Summary of work where the Vishniac instability has been
  identified in hydrodynamic simulations of WBBs. $n_{\rm amb}$,
  $T_{\rm amb}$ and $c_{\rm amb}$ are the ambient number density,
  temperature and sound speed, respectively. Our calculations were
  computed using the value noted for $c_{\rm amb}$.
  The paper references are as follows: SS98 =
\citet{Strickland:1998}, D01 = \citet{Dwarkadas:2001}, F06 =
\citet{Freyer:2006}, N11 = \citet{Ntormousi:2011}, vM12 =
\citet{vanMarle:2012}, K13 = \citet{Krause:2013}. 
}
\begin{tabular}{lllllllll}
\hline
\hline
Paper & $n_{\rm amb}$ & $T_{\rm amb}$ & $c_{\rm amb}$ &
$\dot{M}$ & $v_{\rm wind}$ & $\dot{E}$ & $t_{\rm c}$ & $t_{\rm nr}$ \\
& (${\rm cm^{-3}}$) & (K) & $(\kmps)$ & $(\Msolpyr)$ & $(\kmps)$ &
(${\rm ergs \,s^{-1}}$) & (yr) & (Myr) \\
\hline
SS98 & 10 & $10^{4}$ & 15 & $5\times10^{-5}$ & 2000 &
$6.3\times10^{37}$ & 4000 & 0.17 \\ 
D01  & $\sim 1$ & $\sim 10^{2}$ & 1.5 & $\sim 10^{-7}$ & $\sim 3000$ &
$\approx 2.8\times10^{35}$ & 4200 & 20 \\
F06  & 20 & $8 \times 10^{3}$ & 13.4 & $\sim 3 \times 10^{-7}$ & $\sim
4000$ & $1.5\times10^{36}$ & 810 & 0.025 \\
N11  & 1 & $8 \times 10^{3}$ & 13.4 & \multicolumn{3}{c}{\leftarrowfill \,
  Cluster \rightarrowfill} & 21300 & 0.79 \\
vM12 & 20 & $10^{2}$ & 1.5 & $10^{-6}$ & $2000$ & $1.26\times10^{36}$ &
770 & 9.4 \\
K13  & 10 & $121$ & 1.65 & \multicolumn{3}{c}{\leftarrowfill \, 3 winds +
  SNe \rightarrowfill} & 2300 & 30 \\
\hline
\end{tabular}
\label{tab:literatureresults}
\end{table*}

\subsection{SNR Results}
\label{sec:snr_discussion}

\subsubsection{Numerical and Experimental Investigations of the Vishniac Instability}
The growth of the Vishniac instability during the pressure-driven
snowplough stage of a supernova remnant has been investigated
numerically by \citet{MacLow:1993}, \citet{Blondin:1998} and
\citet{Michaut:2012}. \citet{MacLow:1993} used a 2D spherical polar
grid with $128\times256$ $(r,\theta)$ zones to investigate the
stability of adiabatic blast waves with various values of
$\gamma$. They found that the Vishniac instability saturates for
relatively small amplitudes, but can still produce density variations
of more than a factor of two in the swept-up shell. They also note
that there is only a finite time during which the cold SNR shell is
unstable, because for a typical SNR evolving into a warm ionized ISM
with $n_{\rm 0} < 1 \pcm3$, the Mach number after shell formation is
$< 5$ (while growth of perturbations requires $M \gtsimm 2.6$). They
find that the radial perturbation of the shell is always quite small
($\Delta r/R \approx 0.02$). They do not observe any disruption or
fragmentation of the shell, though if many eigenmodes were included in
the initial perturbation this may have occurred.  They claim that the
instability is stronger in SNRs than in WBBs because of the more rapid
deceleration of SNRs. %see p217.

The Vishniac instability has also been investigated by
\citet{Blondin:1998} in their simulations of the radiative phase in
SNRs. Using 2D simulations, with $10^{4}\times250$ effective
$(r,\theta)$ zones and a realistic cooling function, they find that a
1 per cent perturbation (an $l=128$ spherical harmonic) of an average
ISM density of $n_{\rm H0} = 0.84\pcm3$ leads to shell perturbations
of similar size to the shell thickness.  Increasing the average ISM
density to $n_{\rm H0} = 84\pcm3$ causes the shell to break into a
complicated structure of filaments that extend over the outer 10 per
cent of of the SNR radius. However, in this case the shell is very
thin and the non-linear thin-shell instability
\citep{Vishniac:1994,Blondin:1996} dominates the (initial) behaviour.

\citet{Michaut:2012} performed a numerical simulation of the Vishniac
instability during the pressure-driven snowplough stage of a supernova
remnant. They find that the initial growth of the instability is
quickly damped as the remnant expands, the Mach number of the shell
decreases, and the shell thickens. The remnant is observed to slow
extremely rapidly as it expands into a high pressure ($p/k \sim
10^{7}\,{\rm K}\pcm3$), high temperature ($T\sim 10^{6}$\,K)
environment. It remains to be seen if a simulation with a lower
ambient density and temperature will develop more vigorous Vishniac
instabilities that might lead to the rupture of the shell.

Instabilities and radiative shocks have also been produced with lasers
in high energy laboratory experiments \citep[e.g.][and references
therein]{Edens:2010}. However, the interpretation of the results is
often difficult, and clear evidence of the Vishniac instability has
yet to be presented.

\subsubsection{Observations of Blisters}
The shells of SNRs in their PDS stage show complex filamentary
morphologies and velocity fields. The filaments are likely caused by
ripples in the shock front which are viewed at a range of angles
\citep{Hester:1987}. The source of the ripples could be variations in
the pre-shock gas, though it might also be caused by the Vishniac
instability \citep[see][]{MacLow:1993}.

\citet{Shull:1983} reported turbulent motions in Vela's shell with
speeds up to $30\, \kmps$ and characteristic scales of deformation of
$\approx 0.01$\,pc. Absorption datasets of the Vela SNR indicate that
the velocity is chaotic with high observed line-of-sight velocities
near the edges of the remnant as well as towards its centre
\citep*{Jenkins:1984,Danks:1995}. This is unexpected since an ellipse
in velocity-spatial datasets is expected from coherent expansion of
the shell, and indicates that some process is creating transverse
motions.

To interpret the complicated ultra-violet absorption structure found
by \citet{Jenkins:1984}, \citet{Meaburn:1988} introduced the concept
of blisters on the surface of the shell which form when the shell
ruptures, and which self-seal due to the ablation of material from the
main part of the shell. Echelle spectra of H$\alpha$ and [NII]
emission presented by \citet{Meaburn:1988} revealed the presence of
these blisters, which were found to have a maximum size comparable to
the shell thickness. It is interesting to note that our analytical
calculations are in agreement with this
conclusion. \citet{Meaburn:1990} and \citet{Greidanus:1992} present
data of the SNRs IC\,443 and the Cygnus Loop, respectively, which may
also be consistent with the blister model. In Vela, the Cygnus Loop,
and IC\,443 radial velocity variations of order $\pm 100-200\kmps$
exist. Although this is somewhat higher than the expected flow speed
through the nozzle (see Fig.~\ref{fig:snrmodels}c), the gas will
accelerate further as it blows out into the surrounding medium and
such speeds may be achieveable.

More recently, \citet{Gvaramadze:1998,Gvaramadze:1999} interpret
circular filaments on the face of the Vela SNR as due to the expanding
shells of blisters.  The characteristic size of the blisters is about
$1^{\circ}$, which is a much larger scale relative to the size of the
remnant than Eq.~\ref{eq:rblistermax} predicts. To reconcile this
discrepancy requires that the ruptures in the shell be larger than
expected (since $R_{\rm bmax} \propto \dot{E}_{\rm v}^{1/2}$ in
Eq.~\ref{eq:rblisternr}, a rupture with $10\times$ the radius
increases $\dot{E}_{\rm v}$ by 2 dex and $R_{\rm bmax}$ by 1
dex). This is perhaps not beyond the bounds of possibility considering
that the flow through the rupture should ablate the surrounding shell
and thus act to widen the rupture. A detailed hydrodynamic study of
this process would be useful.

Finally, we note that a series of more than 10 neighbouring loops
together form the so-called Honeycomb nebula in the Large Magellanic
Cloud \citep{Wang:1992}. The loops are of sizes $2-3$\,pc, and the
object is interpreted as forming when a SN shock wave encounters an
intervening sheet of dense, but porous, interstellar gas along our
line of sight \citep{Meaburn:1995,Chu:1995}.  The loops appear as
collimated flows with velocity $\sim 100-200\kmps$ directed towards
the observer \citep{Meaburn:1995}. However, some of the cells show
evidence for self-sealing blisters, which may occur when the magnetic
pressure no longer dominates the gas pressure in the flow
\citep{Redman:1999}. \citet{Meaburn:2010} claim that {\it Chandra}
images are more indicative of emission from the centre of each cell
rather than poorly resolved boundary layers. Since some cells also
show red-shifted velocity spikes \citet{Meaburn:2010} discuss the
possibility of a young SNR in the edge of a giant shell. Again, a
detailed hydrodynamic study of this object would be beneficial.

%%%%%%%%%%%%%%%%%%%%%%%%%%%%%%%%%%%%%%%%%%%%%%%%%%%%%%%%%%%%%%%%%%%%%%%%%%%%%%%
\section{Summary}
\label{sec:conclusions}

This paper investigates the size of blisters resulting from breaks in
the swept-up shell of a wind-blown bubble or a supernova remnant. The
breaks are assumed to occur due to the Vishniac instability. We have
developed a simple analytical model to describe the evolution of such
blisters.

We find that blisters on the surfaces of WBBs are almost always small
in scale relative to the size of the bubble ($R_{\rm bstall}/R \ltsimm
0.2$), with this ratio increasing linearly with time until the shell
is thick enough to prevent any further ruptures. The maximum relative
size of the blisters ($R_{\rm bmax}/R$) scales only as $(v_{\rm
  wind}/c_{\rm 0})^{0.5}$ (cf. Eq.~\ref{eq:rblistermaxrel2}).

Since blowouts are confined and the blisters ``self-seal'' and tend to
be relatively small, shell ruptures cannot be globally important to
WBBs (e.g. they cannot be responsible for the lower than expected
temperatures inside WBBs) except in the most extreme cases. It is
clear, therefore, that truly ``leaky bubbles'' require density
inhomogeneities in the surrounding medium \citep[as envisaged by,
e.g.,][]{Tenorio-Tagle:2006,Harper-Clark:2009} and cannot occur in
smooth media. We note, however, that even when shells run over density
inhomogeneities, leakage is not guaranteed, since the shell can also
re-seal behind it producing a long extended tail in the process
\citep[see][]{Pittard:2011}.

The relative size of blisters on SNRs is even smaller, with blisters
resulting from breaks in the swept-up shell of a SNR in its PDS stage
only growing to radii comparable to the thickness of the cold shell
(Eq.~\ref{eq:rblister_over_dR}), unless the blowout widens the rupture
or the initial rupture is larger than expected. Unexpectedly we find
that the maximum relative size of the blisters on SNRs does {\em not}
depend on key parameters of the model, such as the initial mechanical
energy of the explosion, $E_{\rm 0}$, or the density and temperature
of the ambient ISM, $n_{\rm H0}$ and $T_{\rm 0}$ respectively (see
Eq.~\ref{eq:rblistermax}). The predicted size of the blisters and the
velocity of the gas associated with them appear to be in rough
accordance with observations of SNR shells.

Since the initial expansion of the blisters is faster than that of the
main shell, blisters should affect the observed velocity structure of
SNR shells while they exist. Thus SNR shells should only start to show
a coherent velocity versus radius structure, instead of a messy
veocity structure at all radii, once blisters can no longer form,
which occurs when the shell's Mach number has dropped low enough to
shut off the Vishniac instability. Future hydrodynamical simulations
of this process will be of interest.

The analytical model developed in Sec.~\ref{sec:analytical_model} of
course incorporates many simplifications. For instance, we have
assumed that the WBB/SNR and the shell rupture are essentially static
while the blister grows to its maximum size. We have also assumed that
when a WBB expands into an ionized medium that the ionization front is
not subsequently trapped by the swept-up shell. If this were to occur
it would reduce the ambient sound speed, $c_{\rm 0}$, and thus
increase the shell Mach number and the compression of newly swept-up
gas. Ultimately this may allow the Vishniac instability to continue
operating for a longer time in such scenarios, or allow it to
(re-)start (since $t_{\rm nr}$ increases). The behaviour of the
Vishniac instability during such a transition would be interesting to
see. The instability is also likely to be limited by the presence of
magnetic fields parallel to the shock front which act to limit the
compression ratio of the shell and its bending through magnetic
pressure and tension respectively.

\citet{MacLow:1993} note that the Vishniac instability acts more
strongly in SNRs than in WBBs because of the more rapid decleration of
SNRs. However, the deceleration of SNRs around and after the time of
shell formation is complex \citep[see,
e.g.,][]{Bandiera:2004}. Furthermore, the window of opportunity during
which the Vishniac instability can operate is much smaller in SNRs
than in WBBs because the continued energy injection in WBBs keeps the
Mach number of the shell higher for longer (above the critical value
of $\sim 2.6$ for SNRs and $\sim 3.6$ for WBBs). Thus the Vishniac
instability might actually have a greater effect on the integrity of
the cold shells around WBBs than those around SNRs.

Regardless of how the Vishniac instability creates ruptures, we find
that it is the hotter interior gas which exists behind the cold shell
of WBBs that ultimately allows blisters to develop to a larger
relative size compared to those on SNRs. The hotter interior allows
gas to vent through ruptures at substantially higher speed from WBBs
than from SNRs, and so supplies a larger energy flux with which to
grow a blister for a given size rupture (the latter is related to the
thickness of the cold shell and ultimately to the shell radius).  We
note that this difference between WBB and SNR models may be mitigated
somewhat in WBB models which adopt a lower wind speed, and/or by
mass-loading of the bubble interior, either due to evaporation from
the inside surface of the cold shell \citep{Weaver:1977}, or due to
the ablation or evaporation of embedded clumps
\citep{Pittard:2001a,Pittard:2001b}.

Finally, we may wonder at the necessary conditions for the Vishniac
instability to fragment/rupture a cold shell. It is currently unclear
whether this occurs naturally once the perturbations become much
thicker than the shell, or whether the blowout seen in
Fig.~\ref{fig:sim} required the presence of the additional
structure/perturbations in the flow. We intend to address this
question in a future hydrodynamical investigation.

\section*{Acknowledgements}
JMP would like to thank the referee for a report which led to greater
clarification of some parts of the paper.  JMP gratefully acknowledges
past funding from the Royal Society for a University Research
Fellowship and current funding from the STFC. JMP would also like to
thank Tom Hartquist for introducing him to the \citet{Meaburn:1988}
work on self-sealing bubbles while discussing shell ruptures and
blowouts at a conference in Sexten, Italy, in July 2012.

%\label{lastpage}

%\bibliographystyle{mn2e}
%\bibliography{midrefs}

\label{lastpage}

%, such that the shock radius $\propto t^{s}$ 
%To the shock radius one can add a perturbation radius $\Delta R \propto t^{s}$ 
%with $s = s_{r} + is_{i}$. 

%Observations of blisters: \citet{Gvaramadze:1998} suggest that chaotic motions may form within
%a shell from the development
%of the Richtmyer-Meshkov instability when the SNR collides with a
%pre-existing shell. 

%\citet{Gvaramadze:1998,Gvaramadze:1999} notes that interacting bubbles may alse be present on the shells
%of the SNR G326.3-1.8. 

\end{document}